\documentclass[final,5p,times,twocolumn,authoryear]{elsarticle}

\journal{New Astronomy}

\usepackage{url}
\usepackage[utf8]{inputenc}
\usepackage[T1]{fontenc}
\usepackage{newtxtext,newtxmath}
\usepackage{graphicx}
\usepackage{caption}
\usepackage{subcaption}
\usepackage{float}
\usepackage{hyperref}
\usepackage[normalem]{ulem}
\usepackage[dvipsnames]{xcolor}
\usepackage[normalem]{ulem}
\usepackage{booktabs}
\usepackage{tabularx}
\newcolumntype{C}{>{\centering\arraybackslash}X}

\graphicspath{ {} }

\begin{document}

\begin{frontmatter}

\title{Classification of Eclipsing Binary Light Curves in Gaia DR3:\\ A Machine Learning Approach}

\author[first]{Bedri Keskin}
\author[second,third]{\"Ozg\"ur Ba\c{s}t\"urk}
\affiliation[first]{organization={Ankara University, Graduate School of Natural $\&$ Applied Sciences, Astronomy $\&$ Space Sciences Department, 06100, Tandoğan},
            city={Ankara},
            country={Türkiye}}
\affiliation[second]{organization={Ankara University, Faculty of Science, Astronomy $\&$ Space Sciences Department, 06100, Tandoğan},
            city={Ankara},
            country={Türkiye}}
\affiliation[third]{organization={Ankara University, Astronomy and Space Sciences Research and
Application Center (Kreiken Observatory), Incek Blvd., 06837
Ahlatlıbel},
            city={Ankara},
            country={Türkiye}}

\begin{abstract}
Gaia Data Release 3 (DR3) presents a unique dataset with approximately 2.1 million eclipsing binary star candidates. The unsustainability of manually classifying such a large volume of data has necessitated the development of reliable and scalable automated techniques. In this study, a novel multimodal deep learning model has been developed for the automated classification of approximately 2 million eclipsing binary stars in the Gaia DR3 archive based on their light curve morphologies (EA, EB, EW). The developed architecture simultaneously utilizes a Convolutional Neural Network (CNN) that extracts visual features from light curve images and a Multilayer Perceptron (MLP) that processes geometric model parameters. Noise-free synthetic light curves were used during the training process to ensure the model focuses on geometric shapes. Tests showed that the model achieved an accuracy rate of over 95\% for all classes, exhibiting excellent separation performance, particularly in EA-type systems. As a result of the automated classification performed with the trained model, 40\% of the Gaia DR3 eclipsing binaries were classified as EA, 30\% as EB, and 30\% as EW. This study provides a highly accurate and transferable classification framework for future large-scale sky surveys.
\end{abstract}


\end{frontmatter}

\section{Introduction}
\label{introduction}

Eclipsing binary stars (EBs) are fundamental to modern stellar astrophysics, serving as the primary source for the most precise and model-independent measurements of fundamental stellar parameters, including mass, radius, and effective temperature. The analysis of their light curves allows us to determine the values of these parameters as well as crucial orbital properties, such as the orbital inclinations, that are inaccessible through other methods. EBs can be broadly classified into three main types based on the morphology of their light curves, which are related to the physical configuration of their components: Algol-type (EA), $\beta$-Lyrae type (EB), and W Ursae Majoris-type (EW). Accurate and automated classification of these types is essential because the distinct geometric and physical configurations of the binary components for each type (e.g., detached, semi-detached, or contact systems) directly inform the appropriate stellar evolution models. The publication of Gaia Data Release 3 (DR3) in 2022 provided the scientific community with the most comprehensive survey of the Milky Way to date, containing photometric, astrometric and spectral data for over 1.8 billion objects (\cite{2016AandA...595A...1G} and \cite{2023AandA...674A...1G}). A key focus of the research leveraging this dataset is the identification, classification, and detailed characterization of variable objects. Given the sheer volume of data generated by Gaia and other major photometric surveys such as Kepler \citep{Borucki_2010}, Transiting Exoplanet Survey Satellite (TESS) \citep{Ricker_2015}, PLAnetary Transits and Oscillations of stars (PLATO) \citep{Rauer_2014}, and the upcoming Vera C. Rubin Observatory's Legacy Survey of Space and Time (LSST) \citep{Ivezic_2019}, developing classification methods that are highly robust, scalable, and readily transferable across different survey datasets is a critical requirement for modern time-domain astronomy.

The exponential growth in time-series photometric data from these survey datasets presents a significant challenge: manually classifying millions of variable stars is unsustainable. This has necessitated a shift toward automated classification techniques capable of efficiently processing the immense data volumes delivered by Gaia and similar surveys. A common approach involves characterizing the light curve by extracting a set of diagnostic, numerically defined parameters that describe its morphology, such as the period, amplitude, color variations, and "tailored" diagnostics such the morphology parameter (as defined by \cite{Matijevic_2012}) or coefficients of simple geometric functions or distributions used to fit the light curves \citep{Mowlavi_2023}. This parameterization serves as a vital dimensionality reduction step, transforming the raw time-series data into a compact feature vector. 

Machine Learning (ML) techniques are now the dominant method for analyzing these feature vectors (e.g., \cite{Daza_Perilla_2023}; \cite{Cokina_2021}). Specifically, supervised ML algorithms, such as Random Forests, Support Vector Machines, and Neural Networks, excel at recognizing the subtle, multi-dimensional patterns inherent to the EA, EB, and EW classes. The key advantage of ML lies in its ability to handle the large number of features and the noise inherent in survey data, enabling fast, objective, and high-fidelity classification. \cite{Suveges_2017} classified eclipsing binaries using CALEB, HIPPARCOS, and Kepler datasets with machine learning methods (PCA, LDA, RF, SOM) and successfully applying CALEB-trained models to the Kepler data, where the Random Forest model showed superior performance over the SOM-based classifier. \cite{Bodi_2021} utilized the OGLE dataset, cleaning light curves with DBSCAN and polynomial fitting, then reduced the high-dimensional light curve models to a 2D parameter space using a two-stage Locally Linear Embedding (LLE) process to group stars with similar light curves. \cite{ulas2025} employed CNN-based methods, including pre-trained models like Faster Region-based CNN (Faster R-CNN), You Only Look Once (YOLO), Single Shot Multibox Detector (SSD), and a custom model, trained on synthetic and TESS data to automatically detect pulsation-like structures in the light curves of eclipsing binary stars. However, a significant difficulty remains in ensuring that the training datasets are fully representative and unbiased, and that the resulting classifiers maintain generalizability—that is, the ability to perform equally well when applied to datasets from different surveys (like TESS or LSST) which may have different cadence, noise characteristics, and filter passbands.

The unprecedented scale of the Gaia DR3 release immediately spurred numerous efforts to apply these automated techniques to its data. Among these, the main catalogue compilation work laid the essential foundation for subsequent classification studies. \cite{Eyer_2023} conducted the largest all-sky variability analysis of 1.8 billion objects published in Gaia DR3 using photometric, astrometric and spectral data. A combination of statistical methods and supervised machine learning classifiers were employed to characterize and classify the 10.5 million objects that are identified as variable. Of these, 9.5 million were classified as variable stars and 1 million as active galactic nuclei (AGN)/quasars.

\cite{Gavras_2023} compiled a comprehensive set of objects known from the literature and cross-matched them with Gaia DR3 catalogue. Astrometric and photometric methods were used to find Gaia counterparts for objects in the literature. 152 different catalogues in the literature were cross-matched with Gaia DR3. The resulting catalog contains 7.8 million objects, comprising 112 types of variables, constants, and galaxies. Of these, 1.2 million are non-variable objects, 1.7 million are galaxies, and 4.9 million are variable objects. It contains over a hundred variable classes/subclasses.

\cite{Rimoldini_2023} classified variable objects in the Gaia DR3 archive using eXtreme Gradient Boosting and Random Forest methods. A total of 12.4 million objects were classified, including approximately 9 million variable stars of 22 types, thousands of supernovae, 1 million active galactic nuclei, and 2.5 million galaxies. Eleven variable types were further studied under the name SOS (Single Object Studies).

As a SOS study, \cite{Mowlavi_2023} examined 2,184,477 stars classified as eclipsing variables in the Gaia DR3 archive. The G-band light curves were filtered according to the criteria for eclipsing variables. Three period search methods were applied to the cleaned light curves in the G-band to find the period (P) of variation. One or two Gaussian fits and/or one cosine fit were applied to the G-band light curve of each eclipsing variable candidate. The Gaussian fit was used to model the minimum profiles, while the cosine fit was used to model out-of-eclipse variability such as ellipsoidal-like variations in the light curve. The best combination of the variability period and the geometric model was selected by Bayesian model comparison. The light curves were given a global rank indicating the quality of the selected model. The catalog was restricted to stars with periods greater than 0.2 days. Of the 600,000 cross-matches with the literature, 530,000 have already been classified as eclipsing variables, and 93\% of these had periods consistent with Gaia periods.

We provide the details of the procedure we followed to prepare the Gaia DR3 eclipsing binary
candidates data for classification in Section \ref{sec:data}. Section \ref{sec:MLmodel} explains the multimodal machine learning model and Section \ref{sec:results} discusses the results.

All scripts, Gaia DR3 eclipsing binary
candidates data, results, and graphics are available at Zenodo\footnote{\url{https://doi.org/10.5281/zenodo.18360417}}.

\section{Data preparation}
\label{sec:data}
\begin{figure}[h!] 
    \centering 

    \includegraphics[width=0.8\linewidth]{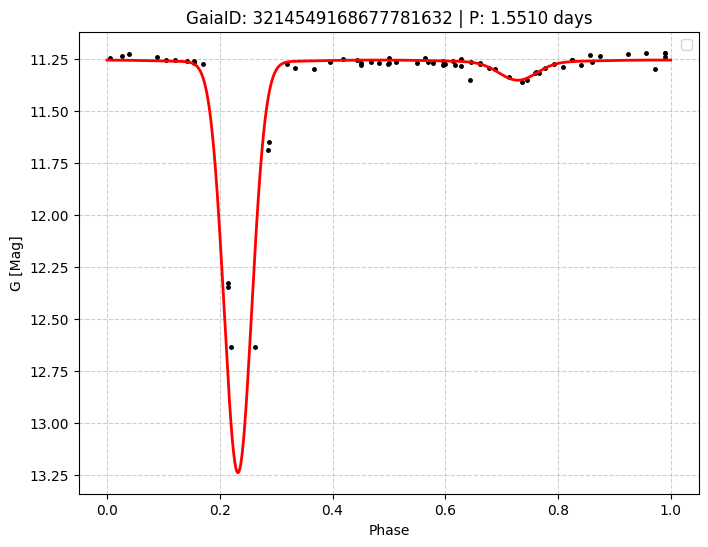}
    \vspace{0.0cm} 

    \includegraphics[width=0.8\linewidth]{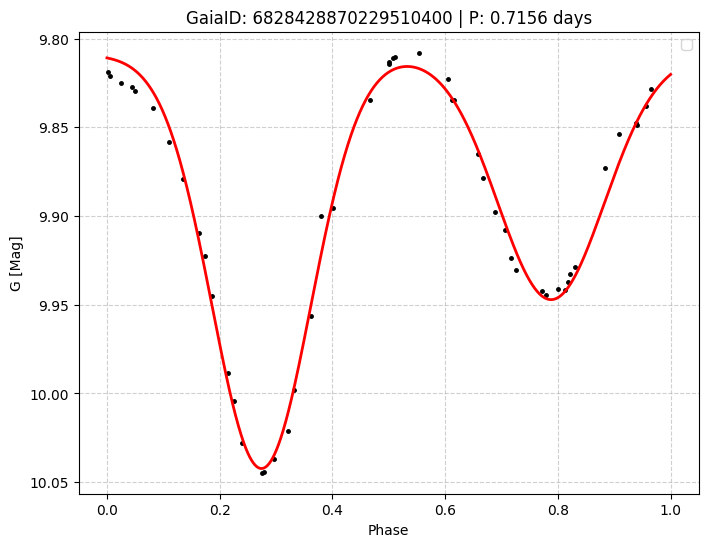}
    \vspace{0.0cm} 

    \includegraphics[width=0.8\linewidth]{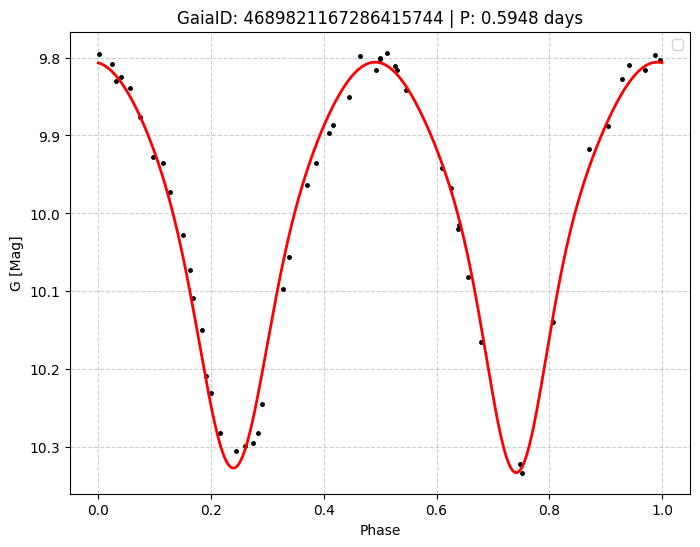}

    \caption{Sample synthetic light curves of EA, EB, EW (from top to bottom) classes obtained based on the geometric model parameters from \cite{Mowlavi_2023} plotted over the phase-folded Gaia G-band light curve data of selected eclipsing binary stars in the Gaia DR3 archive.}
    \label{fig:samples}
\end{figure}

\cite{Mowlavi_2023} modeled Gaia DR3 eclipsing binary candidates in six different ways. All model parameters were downloaded\footnote{\url{https://cdn.gea.esac.esa.int/Gaia/gdr3/Variability/vari_eclipsing_binary/}}. We were able to generate the synthetic light curves of all eclipsing binary candidates (approximately 2.1 million) in Portable Network Graphics (PNG) format (640 × 480 pixels) based on these model parameters to avoid the noise in real data. To demonstrate how well the models fit the light curves, we provide examples of them plotted over phase-folded light curve data in Figure \ref{fig:samples}. More examples can be found in the in the \ref{app:SomeClassificationSamples} and the aforementioned Zenodo repository. For all the stages of machine learning procedure, only the visuals of the synthetic light curves shown in red were used.

The classification of the Gaia DR3 eclipsing binary stars was evaluated separately according to the 6 different model types applied to the light curves by Mowlavi (2023). It was noted that all stars modeled with ``One Gaussian'' model were of the EA (Algol) type, and all objects modeled with an Ellipsoidal model were consistent with the variability expected from  ellipsoidal variables. Therefore, no classification was made for light curves modeled with these two types. For others, classification was performed separately as EA, EB, EW for each model type.

We visually inspected the subsets of light curves from each of these 4 models to be classified and labeled them as EA, EB, and EW for training/validation and testing. The number of manually labeled eclipsing binary stars is given in \ref{app:TrainValTestCounts}. The distributions of eclipsing binary classes in the data sets are ensured to be balanced.

Light curve images were resized to $128 \times 96$ pixels for computation efficiency. They were then converted to numeric data and rescaled to the range [0, 1] at the pixel level. Labels (classes EA, EB, EW) were converted to numeric values using the Label Encoder: 1 for EA, 2 for EB, and 3 for EW. The 80\% of the data was used for training and 20\% for validation according to the well-known Pareto principle. The tabular data was scaled using a normal distribution (standard scaler) with a mean of 0 and a standard deviation of 1. This preprocessing step prevents features with larger numerical scales from dominating the weight updates, thereby facilitating faster convergence of the gradient descent algorithm and ensuring a more stable training process.

\section{The multimodal machine learning model}
\label{sec:MLmodel}

Our classification framework is built upon a multimodal deep learning model that simultaneously processes two distinct types of data: the visual light curve images and the numerical model parameters. Merits of such a multimodal data fusion strategy have been detailed by \cite{Shao_2026}. The architecture consists of two parallel input branches; a Convolutional Neural Network (CNN) for visual feature extraction and a Multilayer Perceptron (MLP) for tabular data which are then concatenated for a final classification.

The visual input branch processes the light curve images. As described in Section \ref{sec:data}, these images were generated from the model parameters and manually labeled. In the preprocessing pipeline, they are resized from their original $640 \times 480$ resolution to $128 \times 96$ pixels with 3 color channels (RGB). Pixel values are normalized to the [0, 1] range. The visual data were processed with CNN layers. First, visual features were extracted with the Conv2D layer. Then, lower-dimensional and meaningful features were extracted with the MaxPooling2D layer. The features were flattened with the Flatten layer and then the features learned in the dense layer were combined.

The numerical input branch processes specific parameters 
derived from model types in \cite{Mowlavi_2023}. The selected features are given in \ref{app:SelectedFeatures}. This tabular data was scaled using a standard scaler to have a mean of 0 and a standard deviation of 1. The numerical data was processed using a multilayer perceptron (MLP). This is followed by two fully connected layers: the first with 32 neurons and the second with 16, both using ReLU activation to produce the final numerical feature vector.

Features extracted from the image and numerical data were combined using the \texttt{concatenate} function. The 16-feature vector from the CNN branch and the 16-feature vector from the MLP branch are concatenated, resulting in a combined 32-feature vector. This combined information from both data types provides a richer and more meaningful representation. The combined vector is passed through an additional dense layer with 32 neurons and rectified linear unit (ReLU) activation. Finally, classification was performed on the combined data. The final output layer is a dense layer with 3 neurons (one for each class: EA, EB, EW) using a 'softmax' activation function to generate class probabilities.

The model was compiled using the ``Adam'' optimizer and ``categorical crossentropy'' as the loss function. Training was conducted with a batch size of 64 for a maximum of 60 epochs. To ensure robustness and prevent overfitting, several callbacks were employed;
\begin{itemize}
\item \textbf{MacroF1 Callback:} A custom callback was implemented to calculate the macro F1-score on the validation set at the end of each epoch. This metric was chosen as the primary performance indicator. It calculates the model's macro F1 score at the end of each epoch and tracks this score on the validation set during training. The F1 score is the harmonic mean of a model's Precision and Recall metrics. The Macro F1 score is the arithmetic mean of the F1 scores calculated separately for each class.

\item \textbf{Model Checkpoint:} This saves the model with the best macro F1 score on the validation set. It also prevents the early stopping strategy from overfitting the model.

\item \textbf{Early Stopping:} If the model's performance on the validation set hasn't improved over time, training is stopped early. This saves unnecessary computational power and time.

\item \textbf{Dropout Layers:} During training, random neurons are shut down, preventing memorization of the model and forcing comprehension.

\item \textbf{L2 Regularization:} Added to dense layers. It prevents the model weights from becoming excessively large.

\item \textbf{ReduceLROnPlateau:} To prevent the model from getting stuck in a local minimum during training, if the validation loss value does not improve for 4 cycles, it halves the learning rate.

\item \textbf{Data Augmentation:} Since the light curve visualizations are very similar to each other, the model is easy to memorize. To differentiate the light curve visualizations, random horizontal flips, right/left or up/down translations, and zooms were applied. Vertical flips should never be performed on light curves.

\end{itemize}

\section{Results and Discussion}
\label{sec:results}
We aimed to develop a robust and automated framework for the classification of eclipsing binaries (EBs) into EA, EB, and EW types, addressing the unsustainable nature of manual classification for large-scale surveys like Gaia DR3. Our data preparation involved a novel approach: rather than training on noisy raw data, we utilized the geometric model parameters from Mowlavi et al. (2023) to generate \textbf{2,184,477} synthetic, noise-free light curve images. We preferred a multimodal deep learning architecture because it allows for the simultaneous processing of visual morphology via a Convolutional Neural Network (CNN) and numerical feature vectors through a Multilayer Perceptron (MLP). This dual-input strategy ensures that the model captures both the intuitive visual "shape" and the precise mathematical characteristics of the stellar minima.

To evaluate the effectiveness of our training strategy and monitor potential issues such as overfitting, the obtained learning curves are given in \ref{app:LearningCurves}. The learning curves illustrate the progression of loss, accuracy, and the Macro-F1 score over training epochs. We observe a rapid convergence within the first five epochs, where the training and validation losses decrease sharply while the accuracy and F1-scores scale upward. The close alignment between the training and validation metrics throughout the process indicates that the model generalizes well to unseen data. This stability is further reinforced by our use of ``Model Checkpoint'' and ``Early Stopping'' callbacks, which ensured that the final weights represented the peak performance on the validation set without succumbing to the noise of over-training.

\subsection{Performance Analysis and Metric Evaluation}
The classification reports on test data is
provided in Tables \ref{tbl:performance_metrics_TWOGAUSSIANS}~-~\ref{tbl:performance_metrics_ONEGAUSSIAN_WITH_ELLIPSOIDAL}.

\begin{table}[ht]
\centering
\begin{tabularx}{\columnwidth}{lCCCC}
\toprule
\textbf{Class} & \textbf{Precision} & \textbf{Recall} & \textbf{F1-score} & \textbf{Support} \\ \midrule
EA & 0.99 & 0.98 & 0.98 & 1019 \\
EB & 0.97 & 0.97 & 0.97 & 1010 \\
EW & 0.98 & 0.98 & 0.98 & 1002 \\ \midrule
Accuracy & & & 0.98 & 3031 \\
Macro avg & 0.98 & 0.98 & 0.98 & 3031 \\
Weighted avg & 0.98 & 0.98 & 0.98 & 3031 \\ \bottomrule
\end{tabularx}
\caption{The classification report of the stars modeled with two Gaussians.}
\label{tbl:performance_metrics_TWOGAUSSIANS}
\end{table}

\begin{table}[ht]
\centering
\begin{tabularx}{\columnwidth}{lCCCC}
\toprule
\textbf{Class} & \textbf{Precision} & \textbf{Recall} & \textbf{F1-score} & \textbf{Support} \\ \midrule
EA & 0.97 & 0.99 & 0.98 & 671 \\
EB & 0.98 & 0.92 & 0.95 & 678 \\
EW & 0.94 & 0.98 & 0.96 & 634 \\ \midrule
Accuracy & & & 0.96 & 1983 \\
Macro avg & 0.96 & 0.96 & 0.96 & 1983 \\
Weighted avg & 0.96 & 0.96 & 0.96 & 1983 \\ \bottomrule
\end{tabularx}
\caption{The classification report of the stars modeled with two Gaussians with ellipsoidal on eclipse 1.}
\label{tbl:performance_metrics_TWOGAUSSIANS_WITH_ELLIPSOIDAL_ON_ECLIPSE1}
\end{table}

\begin{table}[ht]
\centering
\begin{tabularx}{\columnwidth}{lCCCC}
\toprule
\textbf{Class} & \textbf{Precision} & \textbf{Recall} & \textbf{F1-score} & \textbf{Support} \\ \midrule
EA & 0.97 & 0.98 & 0.97 & 529 \\
EB & 0.91 & 0.93 & 0.92 & 535 \\
EW & 0.96 & 0.93 & 0.94 & 544 \\ \midrule
Accuracy & & & 0.95 & 1608 \\
Macro avg & 0.95 & 0.95 & 0.95 & 1608 \\
Weighted avg & 0.95 & 0.95 & 0.95 & 1608 \\ \bottomrule
\end{tabularx}
\caption{The classification report of the stars modeled with two Gaussians with ellipsoidal on eclipse 2.}
\label{tbl:performance_metrics_TWOGAUSSIANS_WITH_ELLIPSOIDAL_ON_ECLIPSE2}
\end{table}

\begin{table}[ht]
\centering
\begin{tabularx}{\columnwidth}{lCCCC}
\toprule
\textbf{Class} & \textbf{Precision} & \textbf{Recall} & \textbf{F1-score} & \textbf{Support} \\ \midrule
EA & 0.99 & 0.99 & 0.99 & 514 \\
EB & 0.97 & 0.94 & 0.96 & 526 \\
EW & 0.95 & 0.98 & 0.97 & 518 \\ \midrule
Accuracy & & & 0.97 & 1558 \\
Macro avg & 0.97 & 0.97 & 0.9 & 1558 \\
Weighted avg & 0.97 & 0.97 & 0.97 & 1558 \\ \bottomrule
\end{tabularx}
\caption{The classification report of the stars modeled with one Gaussian with ellipsoidal.}
\label{tbl:performance_metrics_ONEGAUSSIAN_WITH_ELLIPSOIDAL}
\end{table}

The metrics show an exceptionally high level of performance, with a global accuracy of at least 95\%. These values indicate that the combination of visual CNN features and numerical MLP parameters is highly effective at distinguishing between the EA, EB, and EW classes. Specifically, the high precision suggests a very low false-positive rate, while the consistently high recall across all classes indicates the absence of significant class bias, demonstrating that the model successfully identifies nearly all instances of each binary type. This result validates our multimodal approach as a high-fidelity alternative to traditional single-input classification methods.

The classification reports provide granular assessment over independent test set. The model exhibits near-perfect performance for EA-type stars. This suggests that the detached nature of systems displaying Algol-type light curves creates a distinct signature in both the synthetic light curve images and the numerical model parameters that the multimodal architecture identifies with high confidence. For the EB and EW classes, the metrics remain robust. While these values are slightly lower than those for the EA class, they still represent a high-fidelity classification.

The slight decrease in performance for EB and EW types is the result of the physical and morphological similarities inherent to semi-detached and contact binary systems. In our test set of EB stars and EW stars, the model maintains high precision, indicating that the combination of CNN visual features and MLP numerical features effectively navigates the "fuller" light curve shapes characteristic of these systems. This detailed breakdown confirms that our model is not only accurate on a global scale but is also reliable for populating specific sub-types of light curve morphologies of eclipsing binaries, even when dealing with the subtle transitions between different binary configurations and light curve shapes.

\begin{figure}[H]
    \centering
    \includegraphics[width=0.84\linewidth]{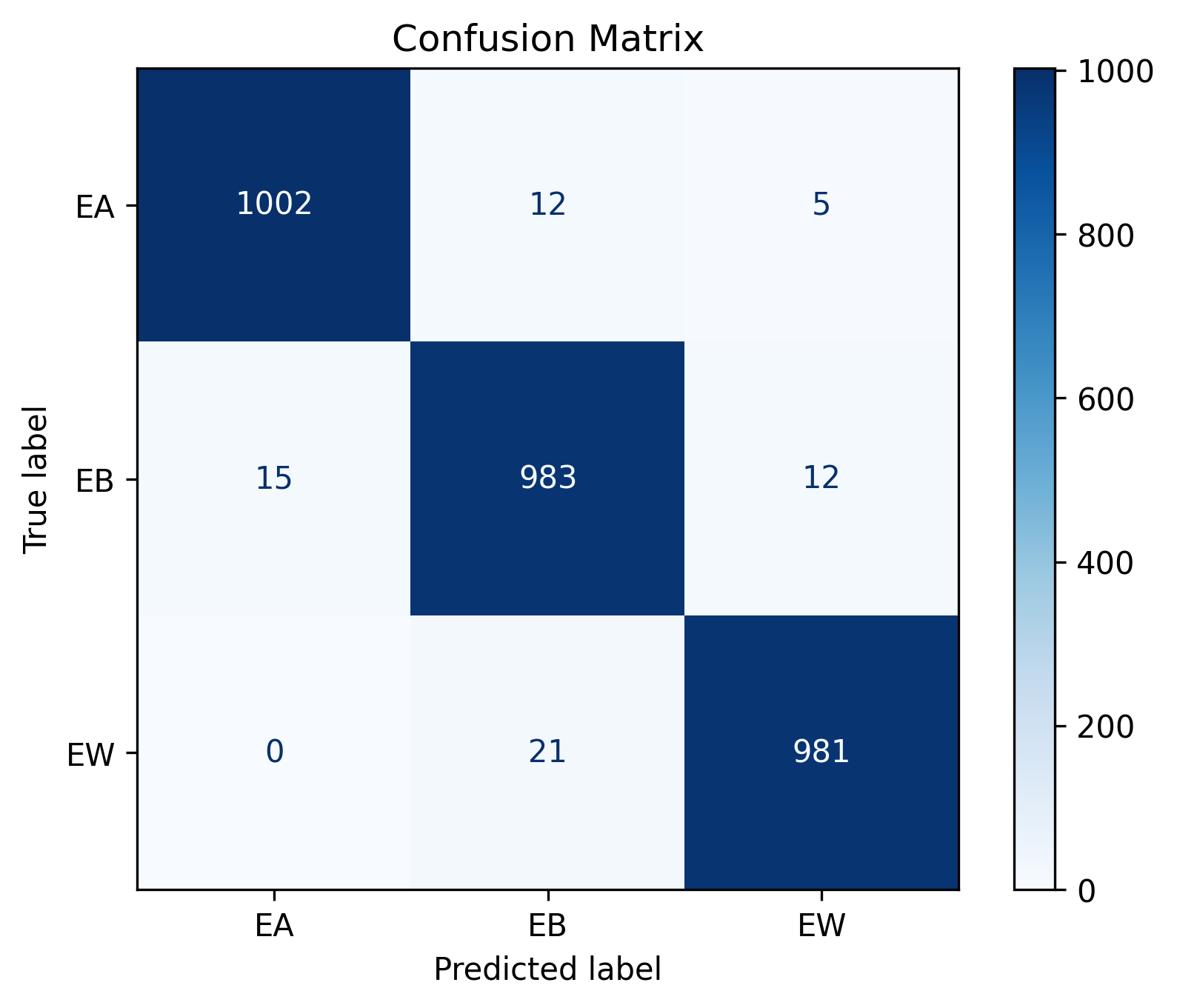}
    \caption{The confusion matrix of the stars modeled with two Gaussians.}
    \label{fig:ConfusionMatrix_TWOGAUSSIANS}
\end{figure}

\begin{figure}[H]
    \centering
    \includegraphics[width=0.83\linewidth]{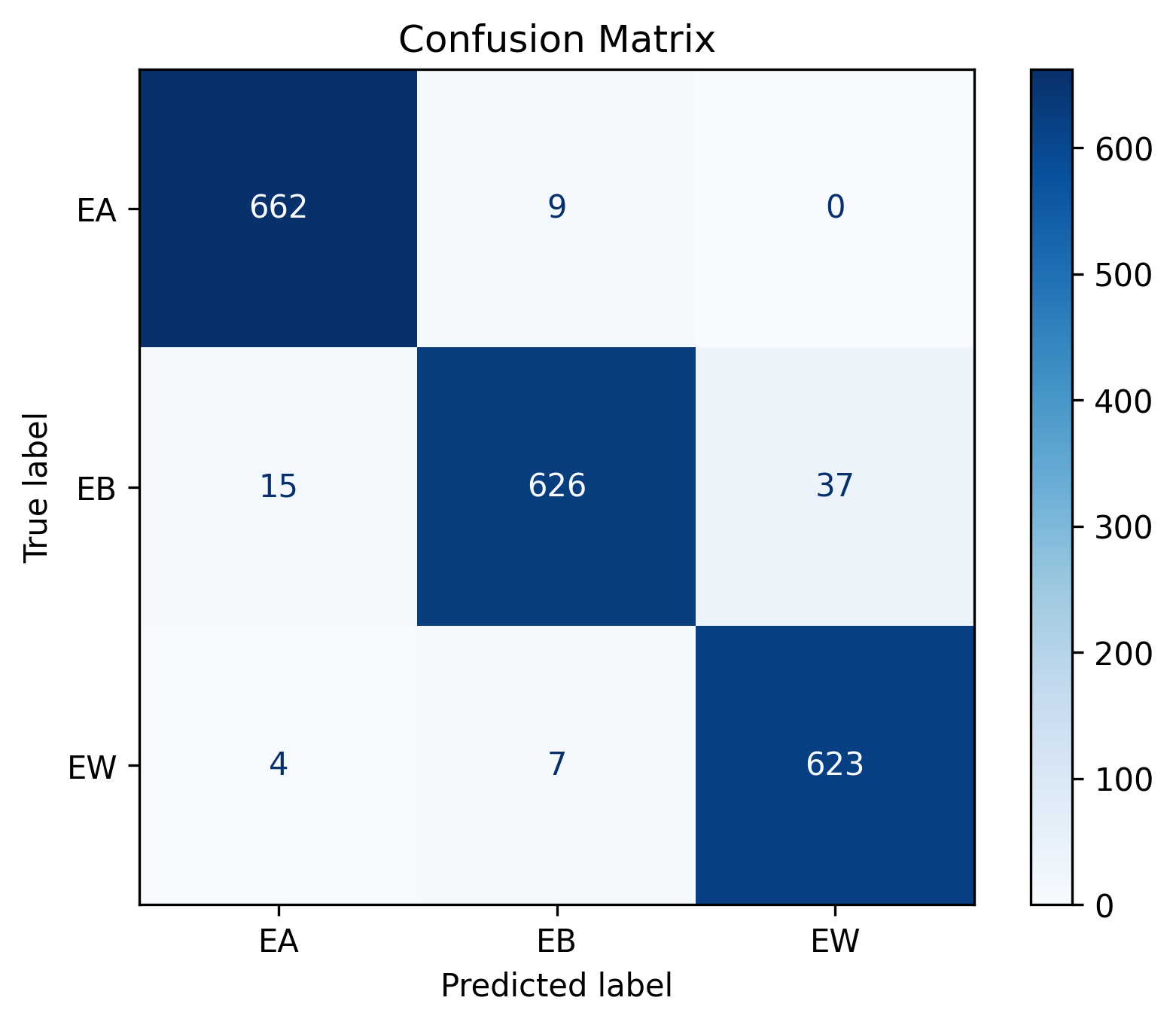}
    \caption{The confusion matrix of the stars modeled with two Gaussians with ellipsoidal on eclipse 1.}
    \label{fig:ConfusionMatrix_TWOGAUSSIANS_WITH_ELLIPSOIDAL_ON_ECLIPSE1}
\end{figure}

\begin{figure}[H]
    \centering
    \includegraphics[width=0.83\linewidth]{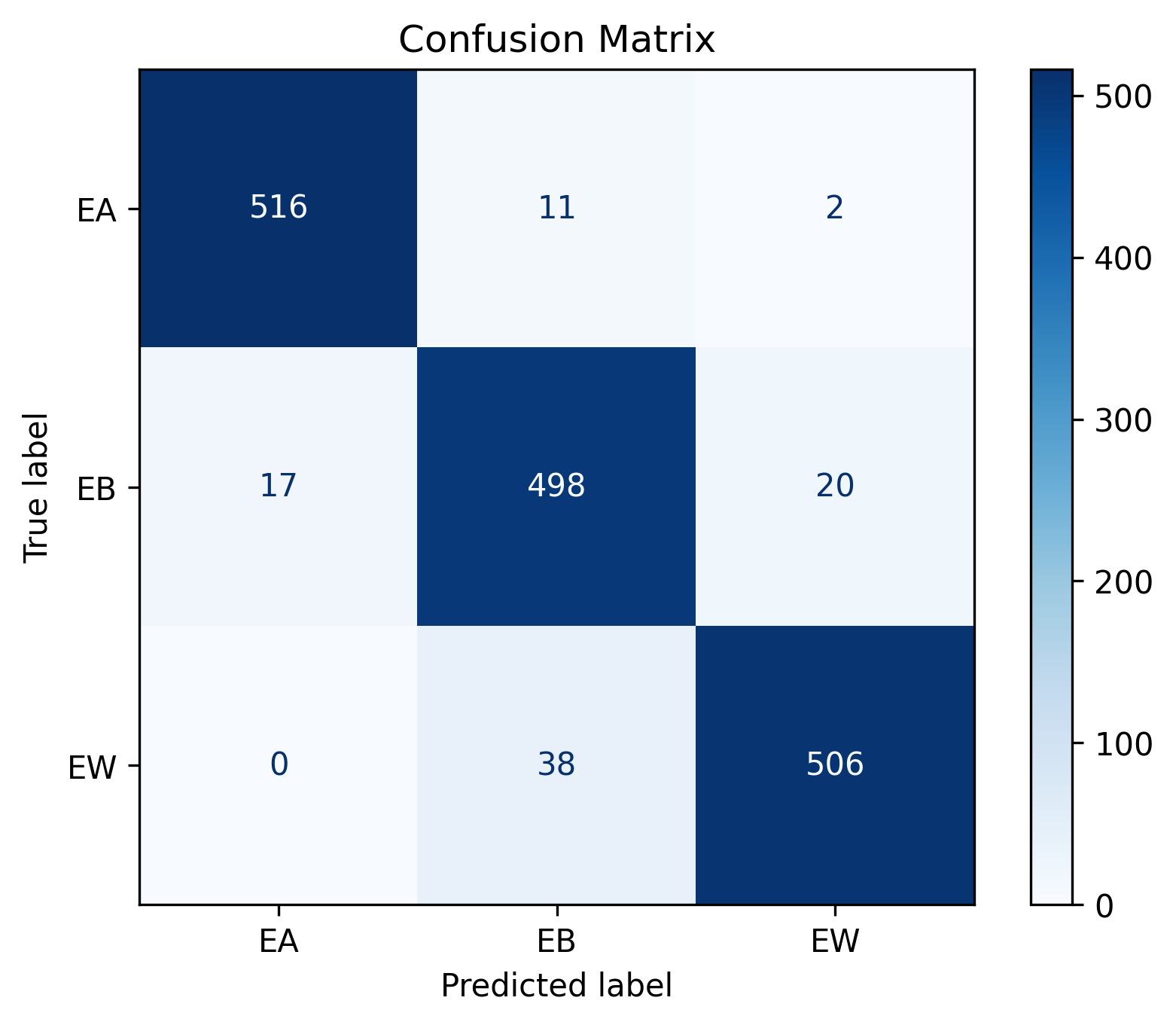}
    \caption{The confusion matrix of the stars modeled with two Gaussians with ellipsoidal on eclipse 2.}
    \label{fig:ConfusionMatrix_TWOGAUSSIANS_WITH_ELLIPSOIDAL_ON_ECLIPSE2}
\end{figure}

\begin{figure}[H]
    \centering
    \includegraphics[width=0.83\linewidth]{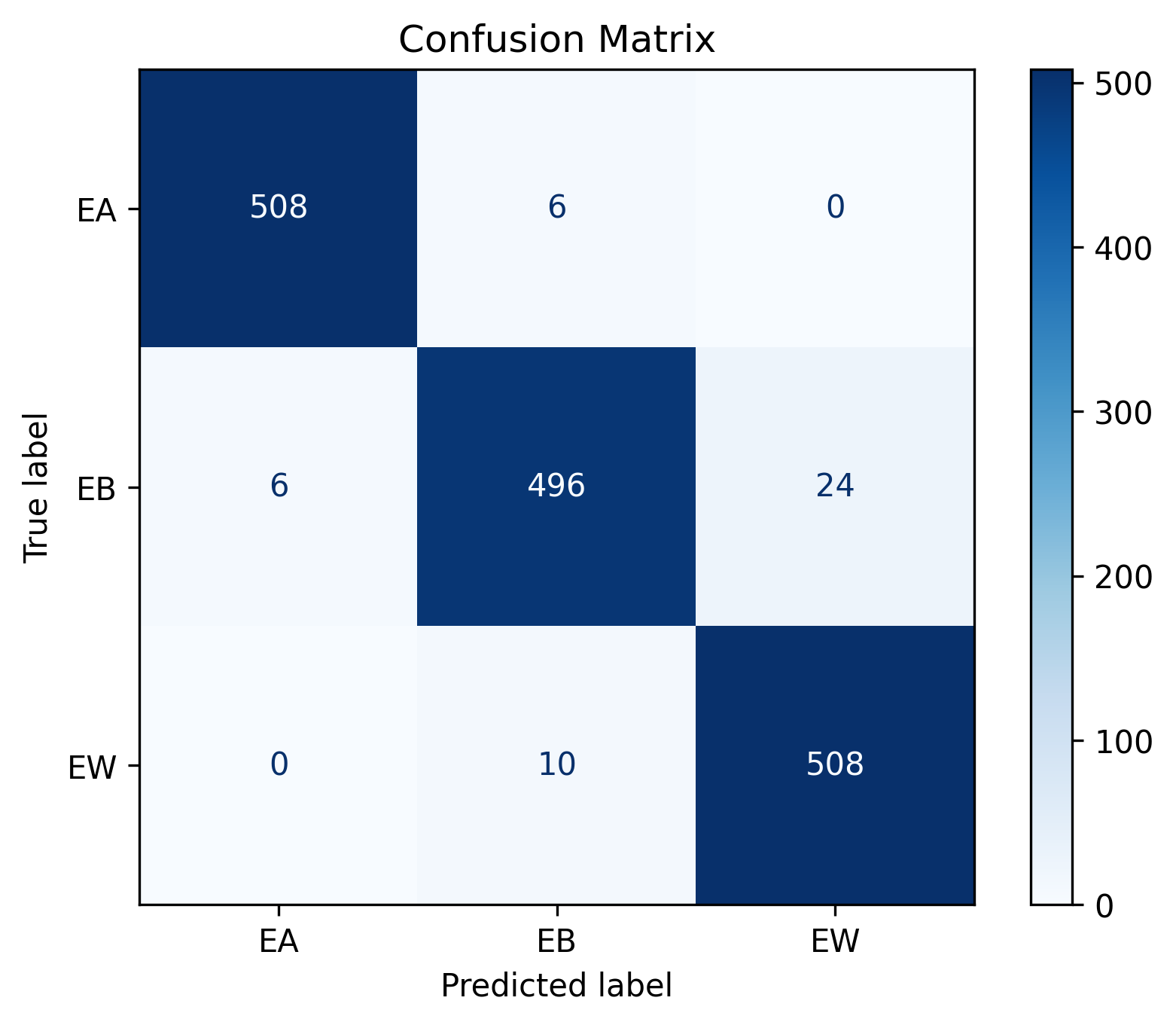}
    \caption{The confusion matrix of the stars modeled with one Gaussian with ellipsoidal.}
    \label{fig:ConfusionMatrix_ONEGAUSSIAN_WITH_ELLIPSOIDAL}
\end{figure}

The confusion matrices are given in Figures \ref{fig:ConfusionMatrix_TWOGAUSSIANS}~-~\ref{fig:ConfusionMatrix_ONEGAUSSIAN_WITH_ELLIPSOIDAL}, which provide a direct visualization of the model's performance on the test datasets.

In these matrices, the rows represent the true labels (manually assigned by the authors), while the columns represent the classes predicted by our multimodal machine learning model. The diagonal elements indicate correct predictions, while the off-diagonal elements reveal the specific nature of any misclassifications. The matrices clearly show that EA systems are identified with nearly perfect accuracy. However, as previously noted, the primary source of classification error, though minimal, occurs between the EB and EW classes. This is a physically anticipated result from the Roche geometries of the systems displaying these two types of light curves. The transition from semi-detached systems displaying mostly EB-type light curves to contact binaries displaying EW-type light curves represents a physical continuum; as the stars evolve and fill their Roche lobes, the sharp distinctions in their light curve morphologies begin to blur. Interpreting the confusion matrix through this astrophysical lens confirms that the model is performing exceptionally well, with its few ``errors'' occurring precisely where the physical definitions of the classes themselves become overlapping and inherently challenging to distinguish.

\subsection{Advantages of the Multimodal Approach}

The success of our framework lies in its multimodal architecture, which processes visual features through a CNN and numerical features through an MLP simultaneously. The use of synthetic light curves generated from model parameters provided a noise-free training set, allowing the CNN to focus purely on the geometric shapes of the minima. This was complemented by the numerical branch, which utilized specific parameters, including depth, sigma, phase of the primary and secondary minima, and amplitude and phase of the cosine component, in order to provide a rigorous mathematical description of the light curve morphology.

This combined feature vector allows for a richer representation of the EB classes than models relying on single-data modes. For instance, while visual data might struggle to differentiate a marginal contact binary from a near-contact semi-detached system, the precise depth and sigma values in the numerical data provide the necessary discriminative power to refine the classification.

Similar multimodal strategies have also been explored in recent astrophysical studies, for instance in quasar classification and redshift estimation from Gaia data \citep{Shi_2026} and in audio-inspired multimodal frameworks for stellar classification \citep{2025FrASS..1259534Z}.

\subsection{Large-Scale Application and Astrophysical Context}

Following successful validation, we applied the model to the broader Gaia DR3 dataset, automatically classifying approximately 2 million stars modeled by \cite{Mowlavi_2023}. The robustness and objectivity of this automated technique address the critical challenge encountered in the classification of timeseries astronomical data. By providing a method that is readily transferable across survey datasets, we offer a framework that can be applied to future high-cadence surveys. Such large-scale, high-fidelity classifications are essential for identification of different light curve types and providing a statistical basis for testing stellar evolution models across different binary configurations that display light curves of these types.

The systematic classification of light curves into EA, EB, and EW types serves as an essential filter for targeted astrophysical research, providing a reliable proxy for the underlying physical configurations. For researchers focused on deriving precise fundamental stellar parameters, the identification of EA-type (Algol) light curves is paramount since these systems typically represent detached configurations where the lack of significant binary interaction signatures, such as strong ellipsoidal variations, allows for the most straightforward and model-independent measurement of masses and radii. Conversely, the classification of EW-type light curves provides a valuable sample for studying the extremes of binary evolution and the influence of third-body dynamics, as the contact systems, that display them, are often found in hierarchical triples where Kozai-Lidov oscillations may have driven them into their current tightly bound \citep{2001ApJ...562.1012E}. Between these two extremes, the EB-type ($\beta$-Lyrae) light curves act as critical morphological proxies for systems in transition, often associated with semi-detached configurations where active mass transfer is occurring. By automating this morphological separation, we enable the efficient construction of specialized catalogs tailored to these distinct scientific objectives.

Out of \textbf{2,111,266} stars, 40,644 were manually classified for training/validation and 8,180 for testing, while all remaining \textbf{2,062,442} stars were automatically classified. The stars have been classified as \textbf{40\% EA}, \textbf{30\% EB}, and \textbf{30\% EW}. Classification details are given in \ref{app:AutoClassification}. Some sample light curves are provided in \ref{app:SomeClassificationSamples}. The archival data used in this study are available in the aforementioned Zenodo repository.

\section*{Acknowledgements}
This work has made use of data from the European Space Agency (ESA) mission
{\it Gaia} (\url{https://www.cosmos.esa.int/gaia}), processed by the {\it Gaia}
Data Processing and Analysis Consortium (DPAC,
\url{https://www.cosmos.esa.int/wgaia/dpac/consortium}). Funding for the DPAC
has been provided by national institutions, in particular the institutions
participating in the {\it Gaia} Multilateral Agreement.

The numerical calculations reported in this paper were fully/partially performed at TUBITAK ULAKBIM, High Performance and Grid Computing Center (TRUBA resources).

\clearpage
\onecolumn 
\raggedbottom
\appendix

\section{Number of manually labeled eclipsing binary stars}
\label{app:TrainValTestCounts}

\begin{table}[ht]
\begin{tabular}{lcc}
\toprule
\textbf{Model Type} & \textbf{Train/Val} & \textbf{Test} \\ \midrule
TWOGAUSSIANS & 15.094 & 3.031 \\
TWOGAUSSIANS\_WITH\_ELLIPSOIDAL\_ON\_ECLIPSE1 & 10.032 & 1.983 \\
TWOGAUSSIANS\_WITH\_ELLIPSOIDAL\_ON\_ECLIPSE2 & 7.981 & 1.608 \\
ONEGAUSSIAN\_WITH\_ELLIPSOIDAL & 7.537 & 1.558 \\ \midrule
\textbf{Total} & \textbf{40.644} & \textbf{8.180} \\ \bottomrule
\end{tabular}
\end{table}

\vspace{2em}

\section{Selected features}
\label{app:SelectedFeatures}

\begin{table}[ht]
\centering
\begin{tabularx}{\textwidth}{lX} 
\toprule
\textbf{Name} & \textbf{Description} \\ \midrule
frequency & Orbital frequency of the EB \\
geom\_model\_reference\_level & Magnitude reference level of geometric model \\
geom\_model\_gaussian1\_phase & Phase of Gaussian 1 \\
geom\_model\_gaussian1\_sigma & Width (standard deviation, in phase) of Gaussian 1 \\
geom\_model\_gaussian1\_depth & Depth of Gaussian 1 \\
geom\_model\_gaussian2\_phase & Phase of Gaussian 2 \\
geom\_model\_gaussian2\_sigma & Width (standard deviation, in phase) of Gaussian 2 \\
geom\_model\_gaussian2\_depth & Depth of Gaussian 2 \\ 
geom\_model\_cosine\_half\_period\_amplitude & Amplitude (half peak-to-peak) of the cosine component
with half the period of the geometric model \\
geom\_model\_cosine\_half\_period\_phase & Phase of the cosine component
with half the period of the geometric model \\ \bottomrule
\end{tabularx}
\end{table}

\section{Learning curves}
\label{app:LearningCurves}

\begin{figure}[H]
    \centering
    \includegraphics[width=1.0\linewidth]{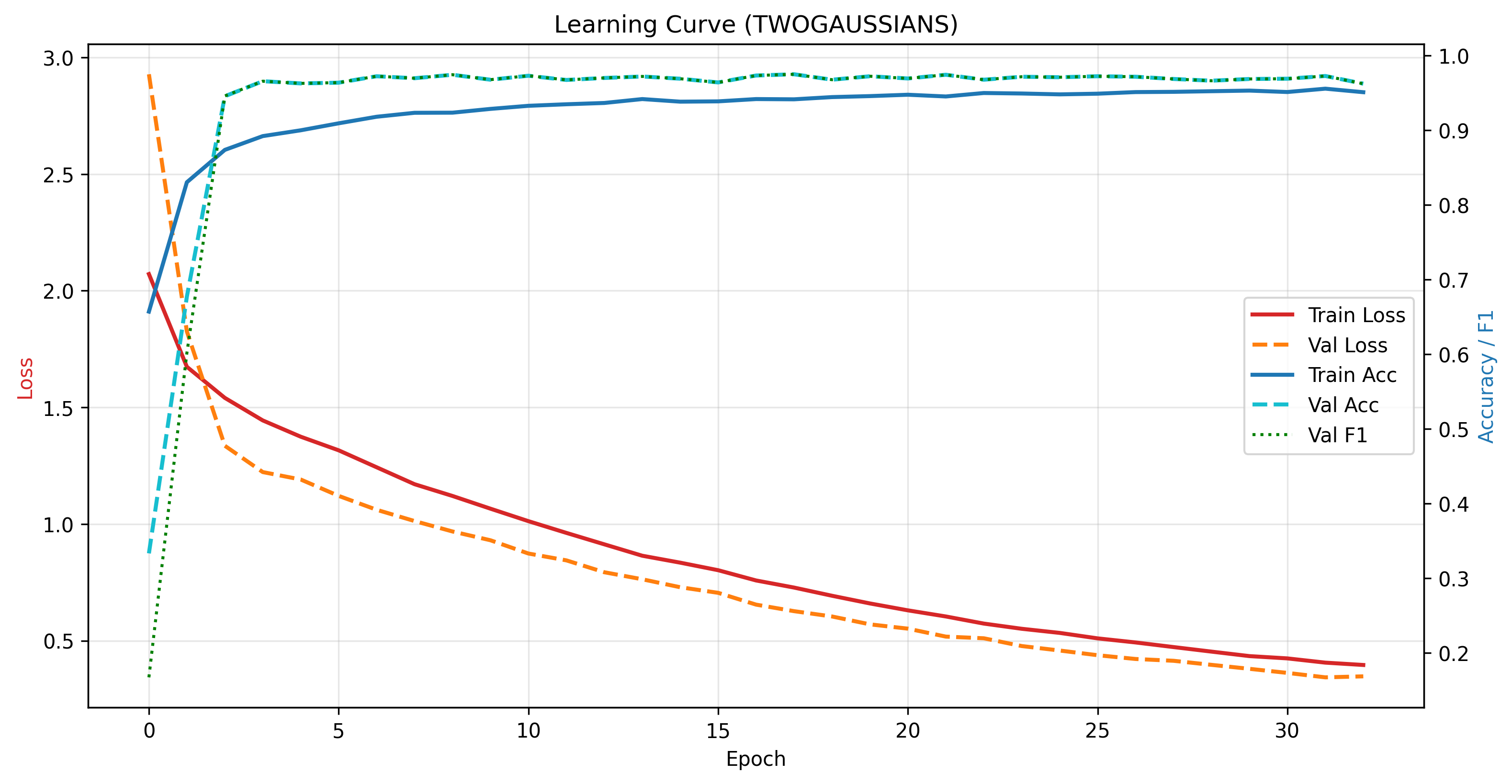}
    \caption{Learning curve of the training of the stars modeled with two Gaussians.}
\end{figure}

\begin{figure}[H]
    \centering
    \includegraphics[width=1.0\linewidth]{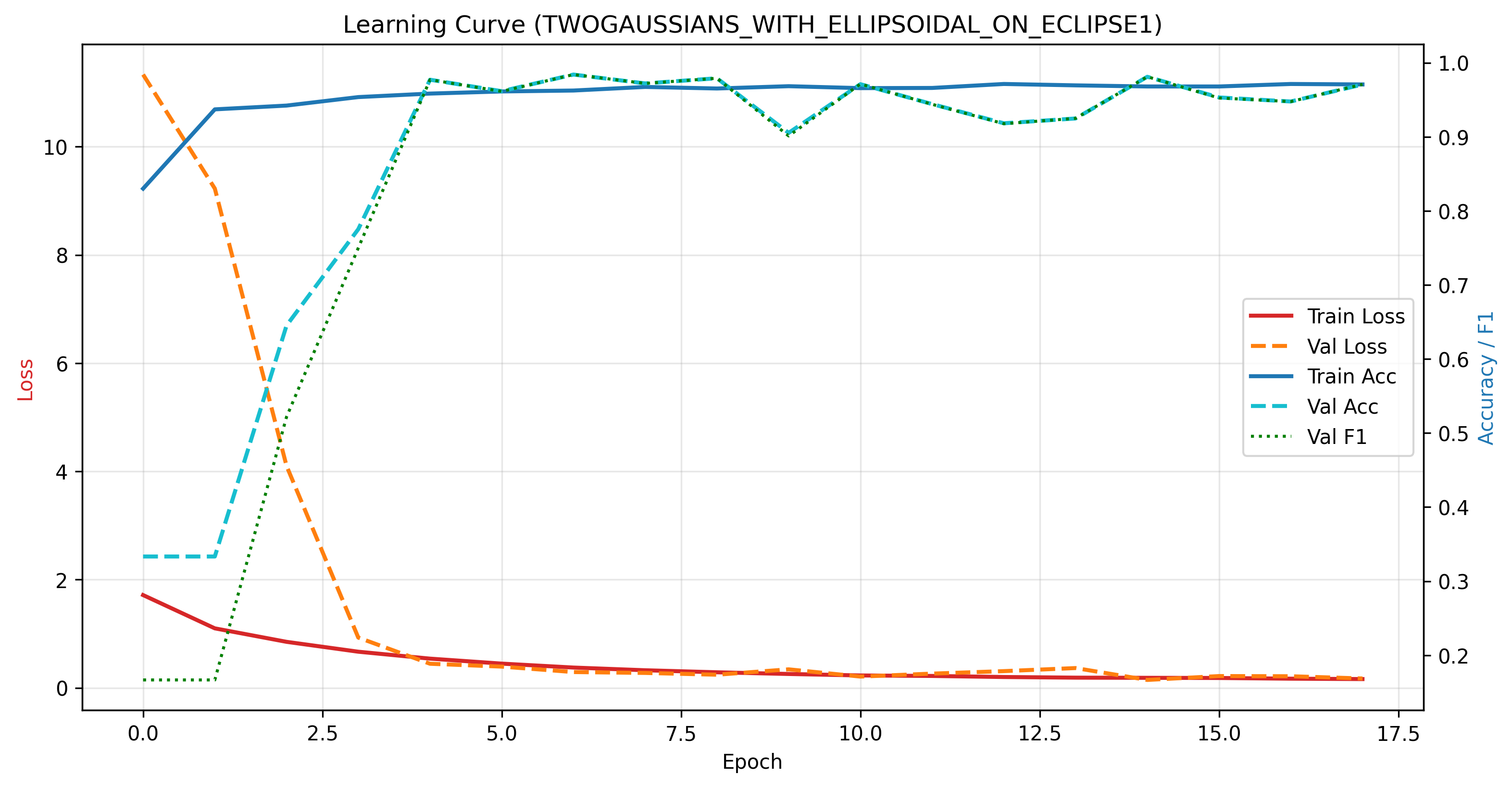}
    \caption{Learning curve of the training of the stars modeled with two Gaussians with ellipsoidal on eclipse 1.}
\end{figure}

\begin{figure}[H]
    \centering
    \includegraphics[width=1.0\linewidth]{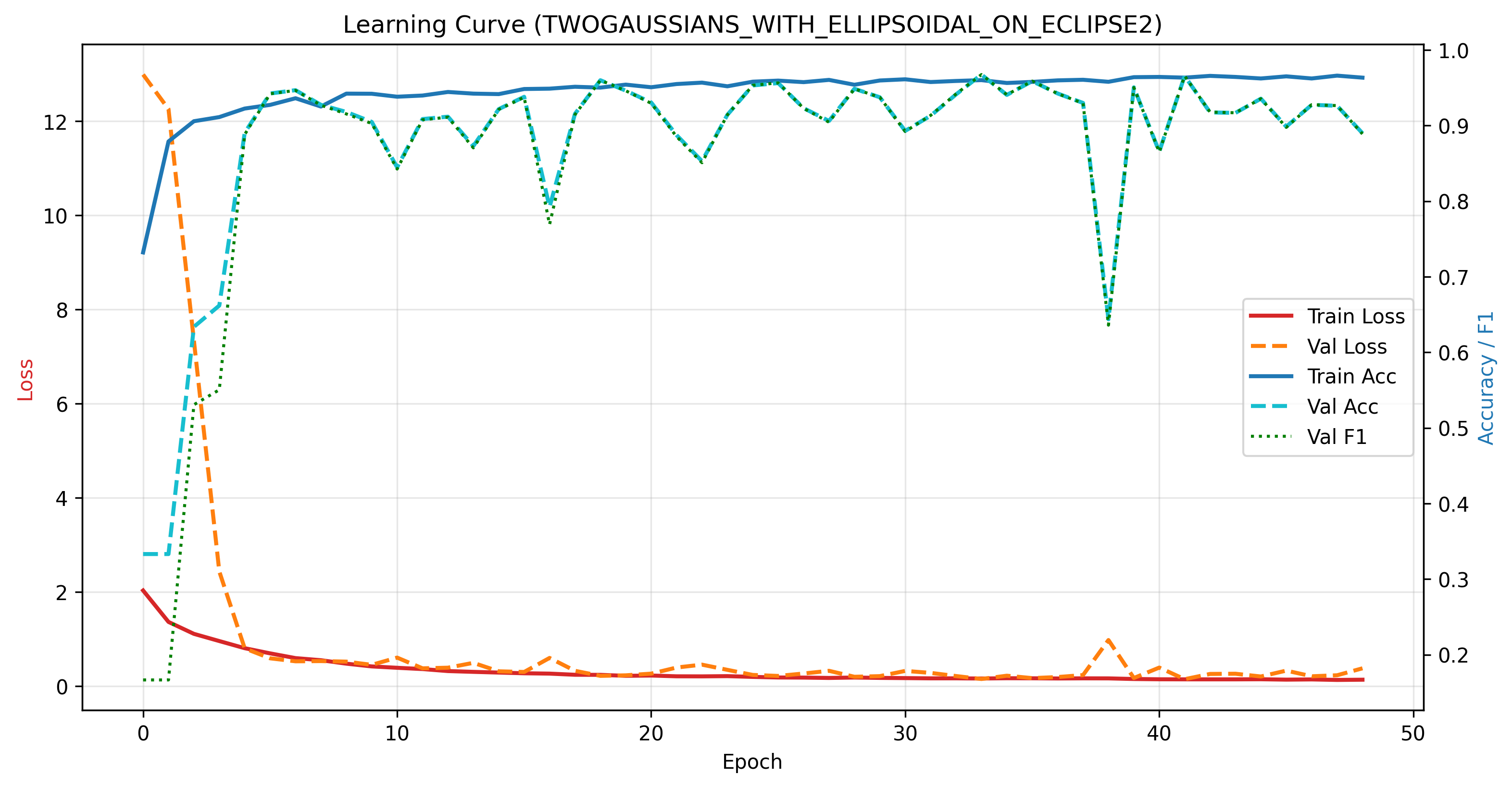}
    \caption{Learning curve of the training of the stars modeled with two Gaussians with ellipsoidal on eclipse 2.}
\end{figure}

\begin{figure}[H]
    \centering
    \includegraphics[width=1.0\linewidth]{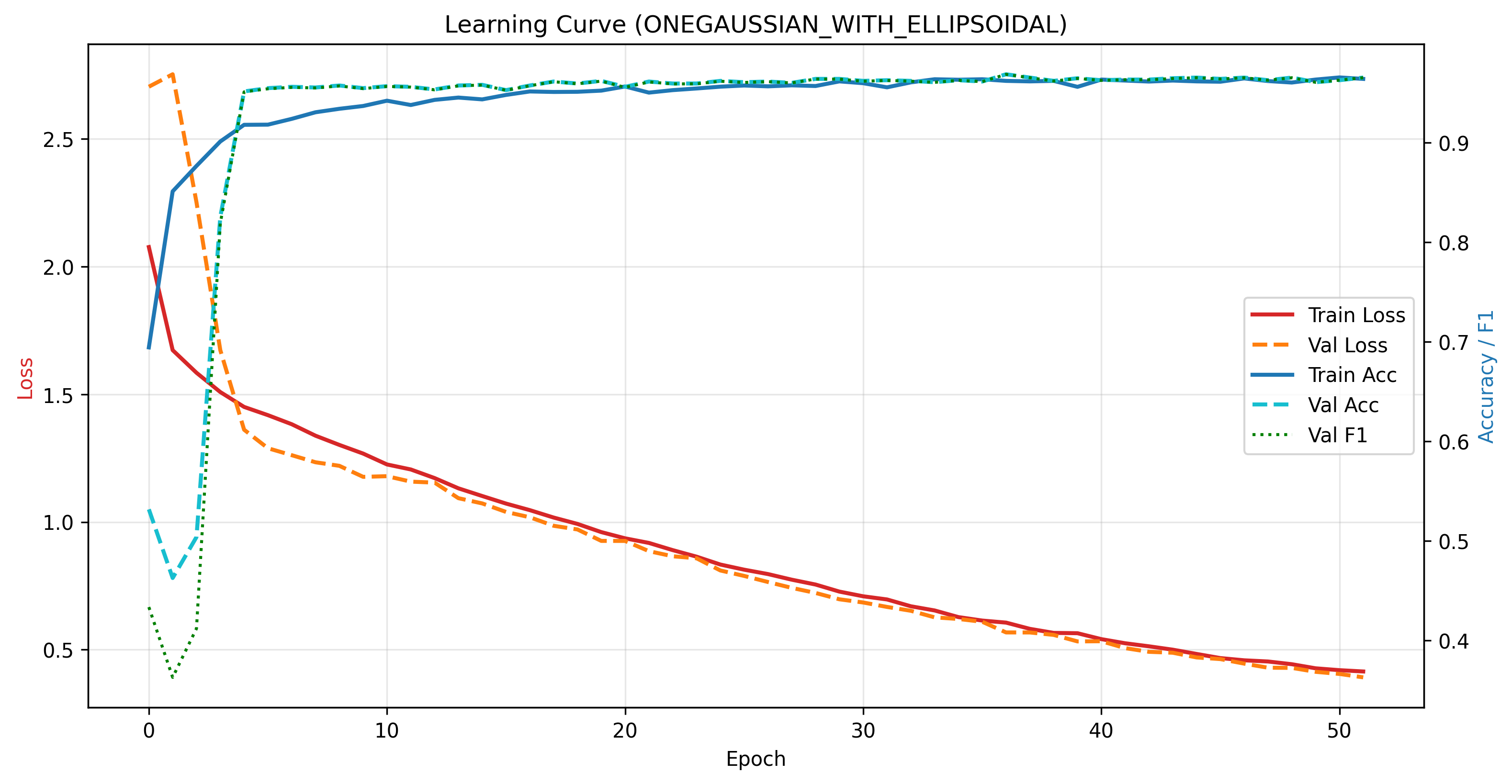}
    \caption{Learning curve of the training of the stars modeled with one Gaussian with ellipsoidal.}
\end{figure}

\section{Numerical distribution of eclipsing binaries automatically classified according to model types}
\label{app:AutoClassification}

\begin{center} 
    \small 
    \begin{tabularx}{\textwidth}{lCCCC} 
        \toprule
        \textbf{Model Type} & \textbf{EA} & \textbf{EB} & \textbf{EW} & \textbf{Total} \\ \midrule
        TWOGAUSSIANS & 703,865 (45\%) & 446,653 (28\%) & 419,283 (27\%) & 1,569,801 \\
        TWOGAUSSIANS\_WITH\_ELLIPSOIDAL\_ON\_ECLIPSE1 & 88,235 (23\%) & 120,673 (32\%) & 168,802 (45\%) & 377,710 \\
        TWOGAUSSIANS\_WITH\_ELLIPSOIDAL\_ON\_ECLIPSE2 & 22,784 (30\%) & 28,834 (38\%) & 24,193 (32\%) & 75,811 \\
        ONEGAUSSIAN\_WITH\_ELLIPSOIDAL & 8,294 (21\%) & 25,014 (64\%) & 5,812 (15\%) & 39,120 \\ \midrule
        \textbf{Total number of automatically classified stars} & \textbf{823,178 (40\%)} & \textbf{621,174 (30\%)} & \textbf{618,090 (30\%)} & \textbf{2,062,442} \\ \bottomrule
    \end{tabularx}
    \label{tbl:AutoClassification}
\end{center}

\vspace{2em}

\section{Some classification samples}
\label{app:SomeClassificationSamples}

\subsection{EA samples}
\label{fig:EAsamples}
\begin{center}
    \begin{minipage}{0.48\textwidth}
        \includegraphics[width=\linewidth]{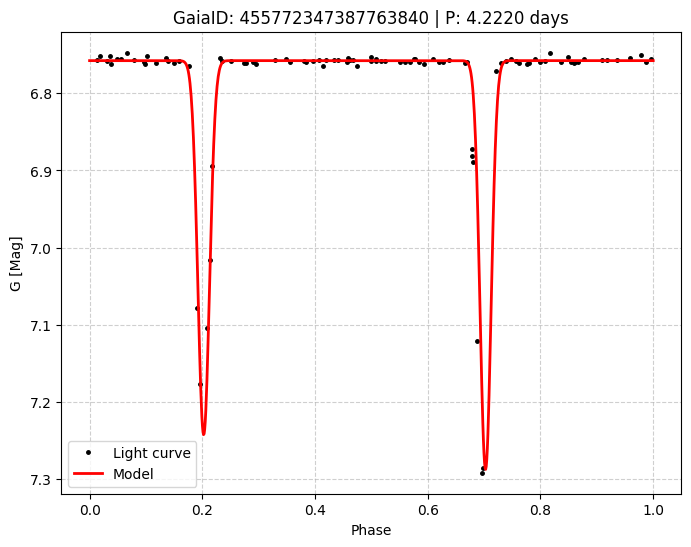}
    \end{minipage} \hfill
    \begin{minipage}{0.48\textwidth}
        \includegraphics[width=\linewidth]{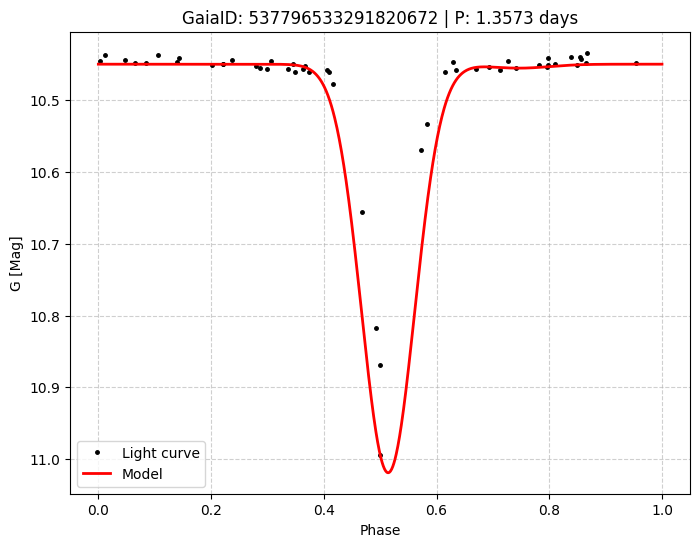}
    \end{minipage}

    \vspace{0.5em}

    \begin{minipage}{0.48\textwidth}
        \includegraphics[width=\linewidth]{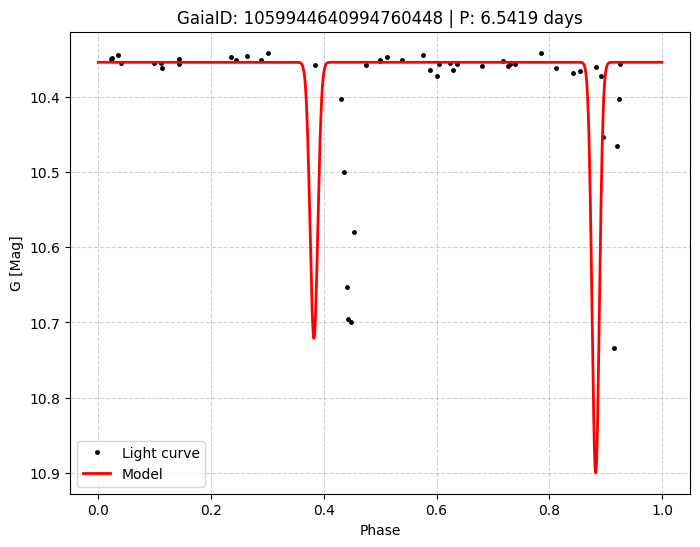}
    \end{minipage} \hfill
    \begin{minipage}{0.48\textwidth}
        \includegraphics[width=\linewidth]{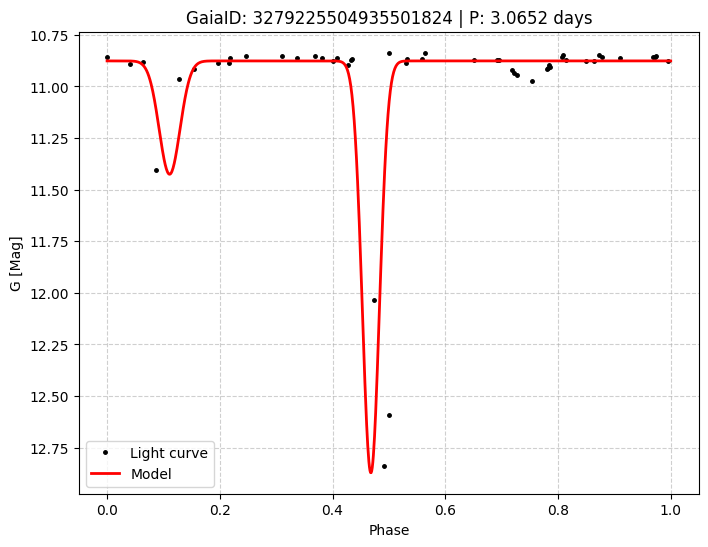}
    \end{minipage}

    \vspace{0.5em}

    \begin{minipage}{0.48\textwidth}
        \includegraphics[width=\linewidth]{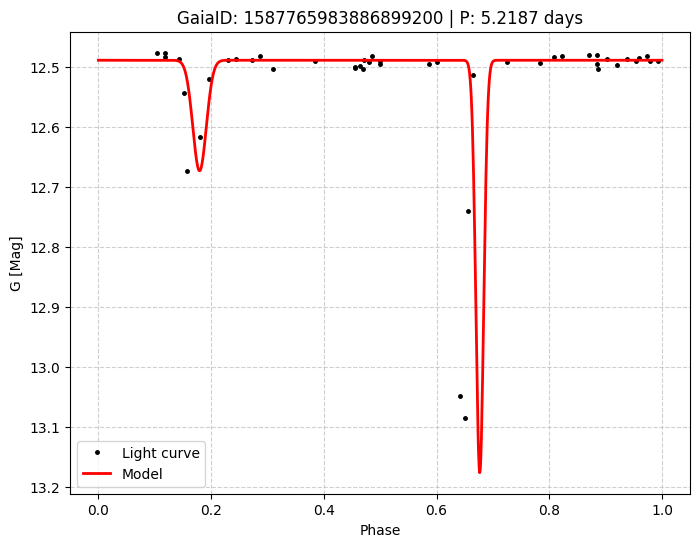}
    \end{minipage} \hfill
    \begin{minipage}{0.48\textwidth}
        \includegraphics[width=\linewidth]{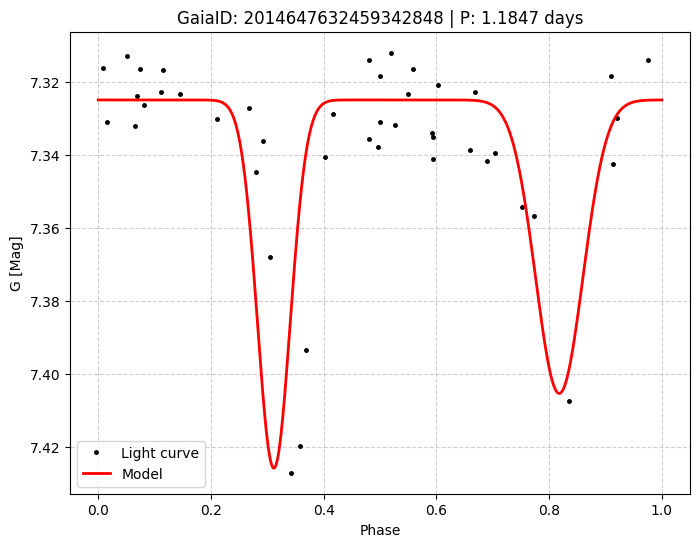}
    \end{minipage}

    \vspace{0.5em}

    \begin{minipage}{0.48\textwidth}
        \includegraphics[width=\linewidth]{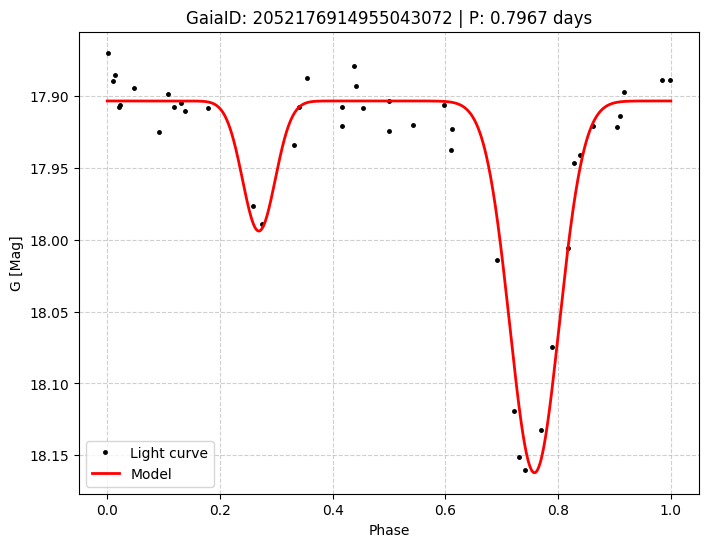}
    \end{minipage} \hfill
    \begin{minipage}{0.48\textwidth}
        \includegraphics[width=\linewidth]{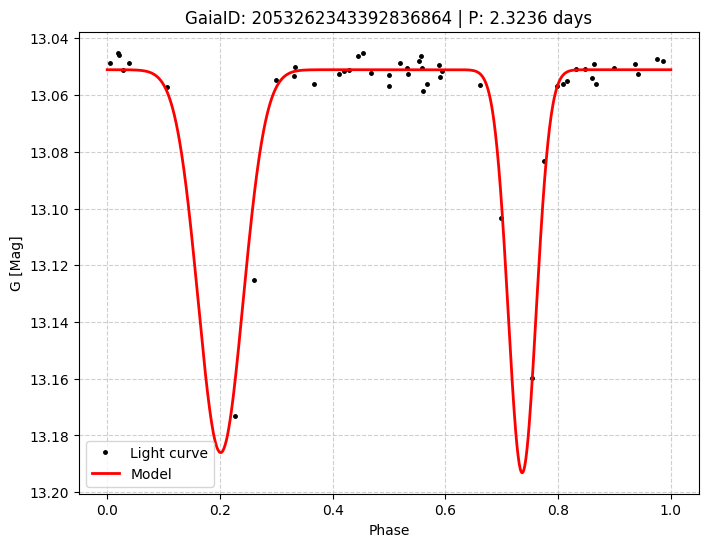}
    \end{minipage}

    \vspace{0.5em}

    \begin{minipage}{0.48\textwidth}
        \includegraphics[width=\linewidth]{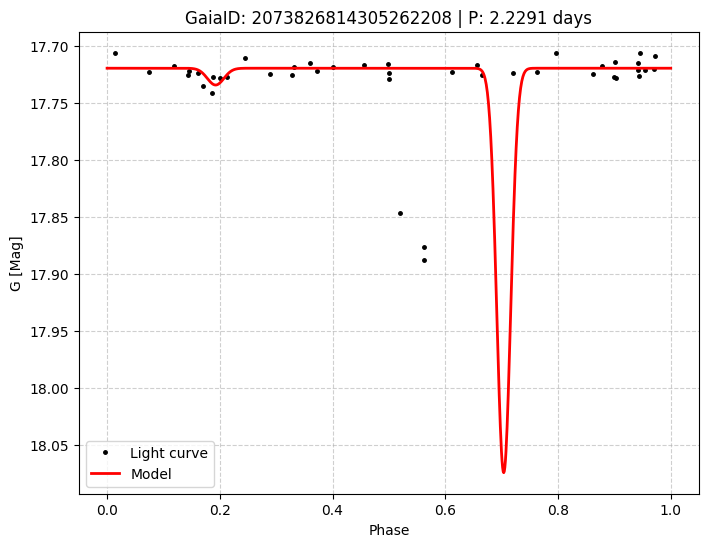}
    \end{minipage} \hfill
    \begin{minipage}{0.48\textwidth}
        \includegraphics[width=\linewidth]{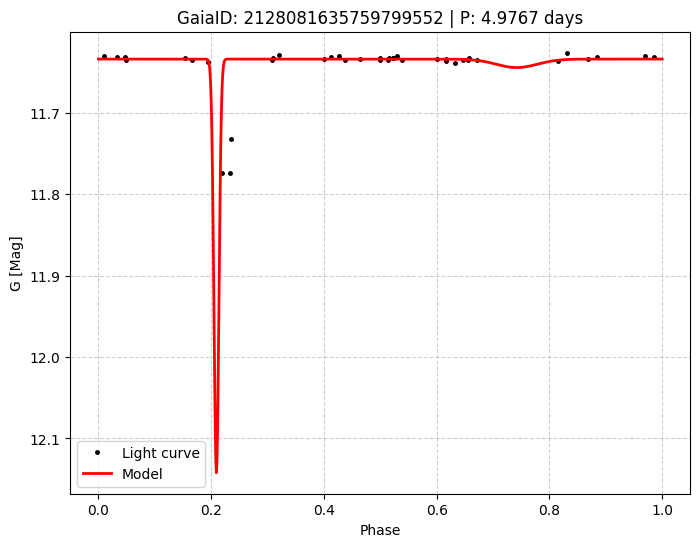}
    \end{minipage}






\end{center}

\subsection{EB samples}
\label{fig:EBsamples}
\begin{center}
    \begin{minipage}{0.48\textwidth}
        \includegraphics[width=\linewidth]{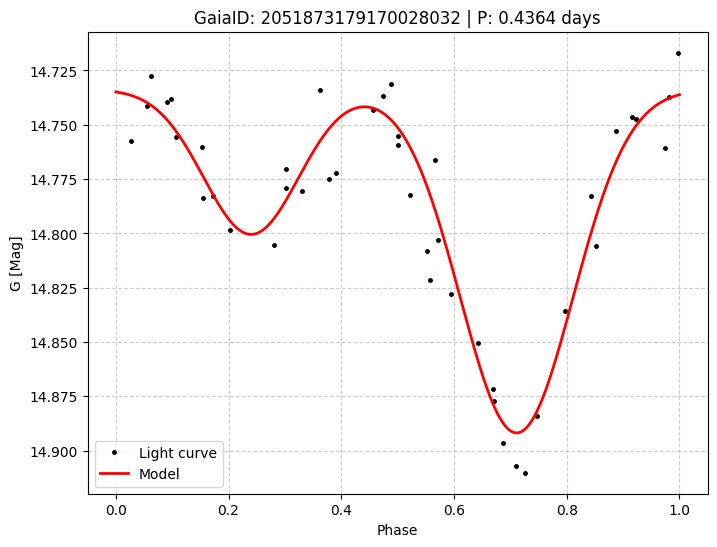}
    \end{minipage} \hfill
    \begin{minipage}{0.48\textwidth}
        \includegraphics[width=\linewidth]{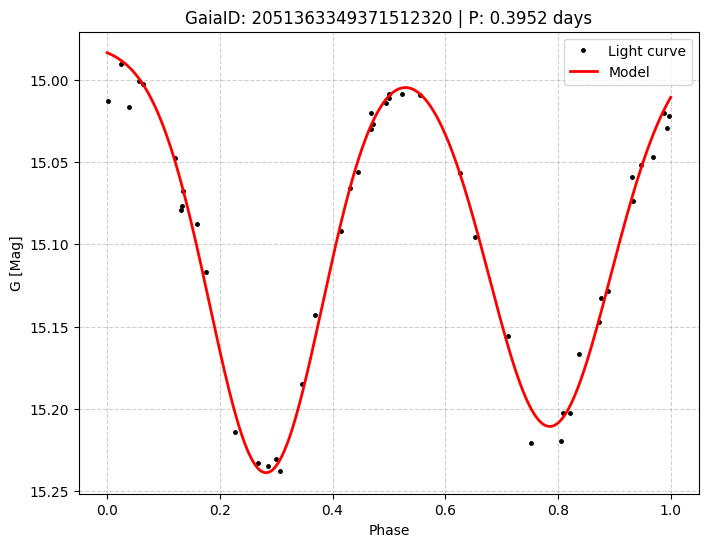}
    \end{minipage}

    \vspace{0.5em}

    \begin{minipage}{0.48\textwidth}
        \includegraphics[width=\linewidth]{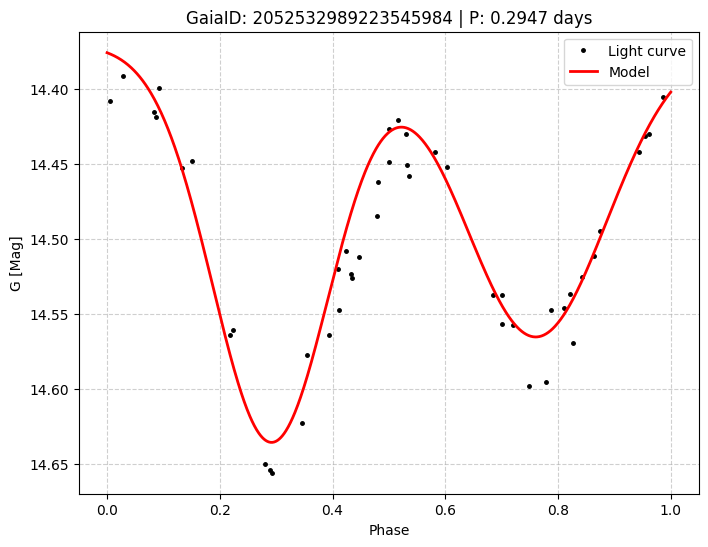}
    \end{minipage} \hfill
    \begin{minipage}{0.48\textwidth}
        \includegraphics[width=\linewidth]{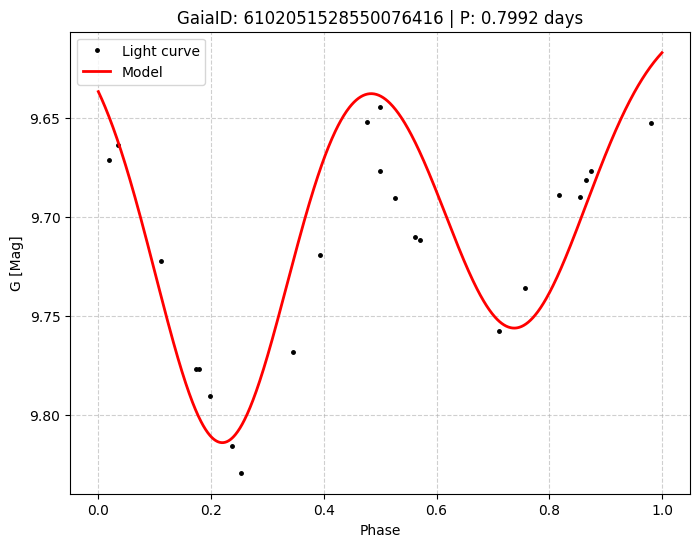}
    \end{minipage}

    \vspace{0.5em}

    \begin{minipage}{0.48\textwidth}
        \includegraphics[width=\linewidth]{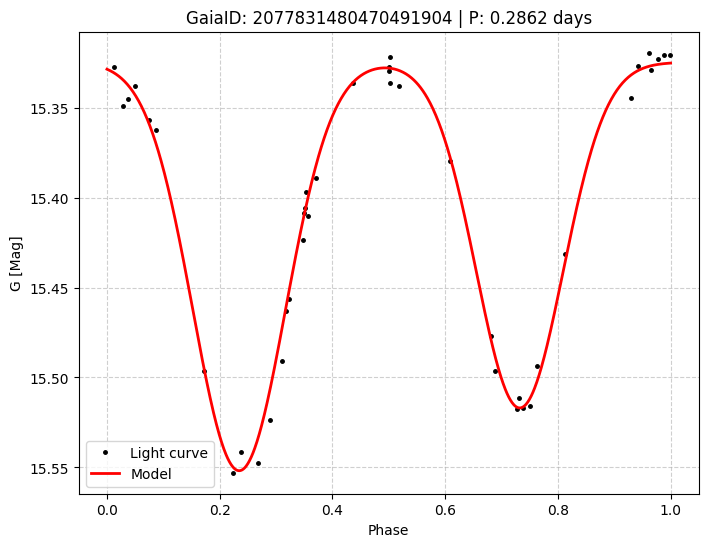}
    \end{minipage} \hfill
    \begin{minipage}{0.48\textwidth}
        \includegraphics[width=\linewidth]{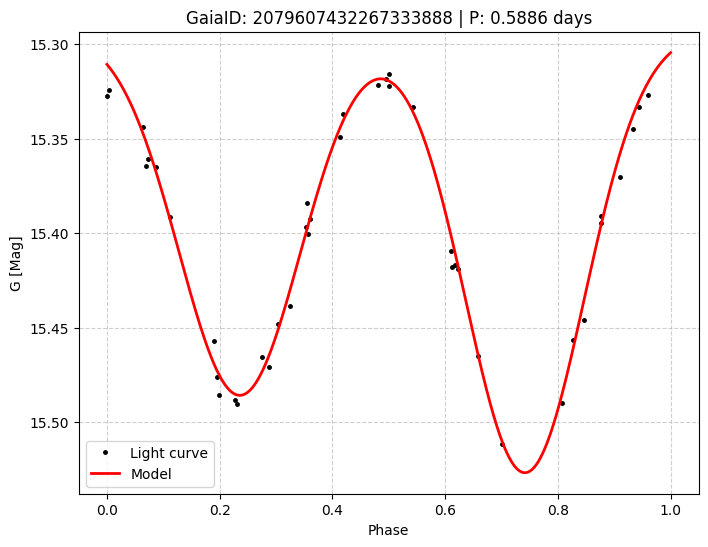}
    \end{minipage}

    \vspace{0.5em}

    \begin{minipage}{0.48\textwidth}
        \includegraphics[width=\linewidth]{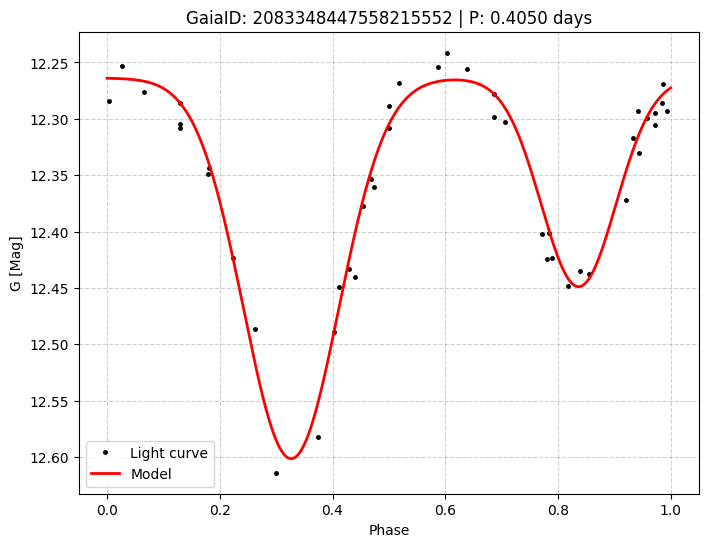}
    \end{minipage} \hfill
    \begin{minipage}{0.48\textwidth}
        \includegraphics[width=\linewidth]{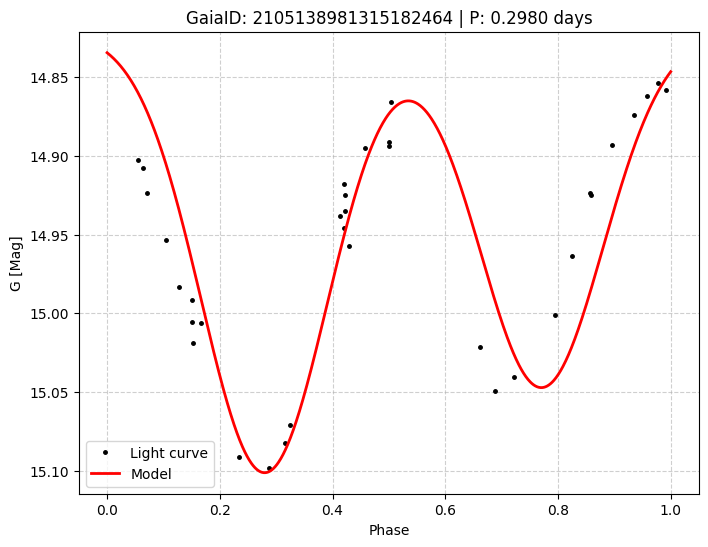}
    \end{minipage}

    \vspace{0.5em}

    \begin{minipage}{0.48\textwidth}
        \includegraphics[width=\linewidth]{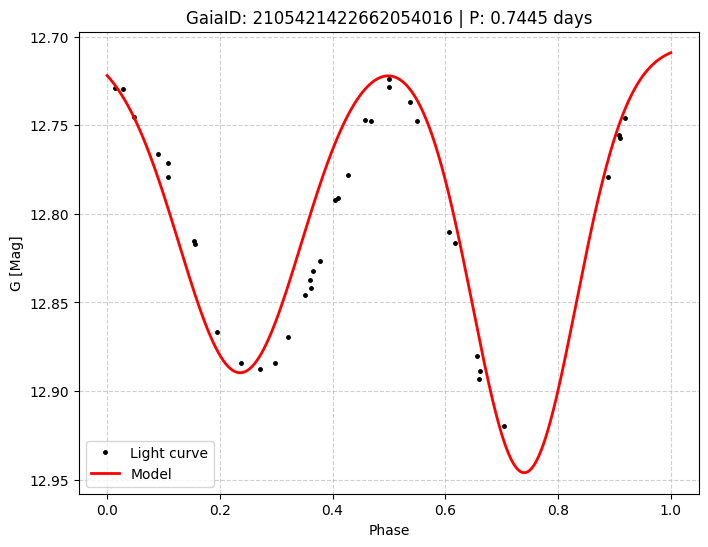}
    \end{minipage} \hfill
    \begin{minipage}{0.48\textwidth}
        \includegraphics[width=\linewidth]{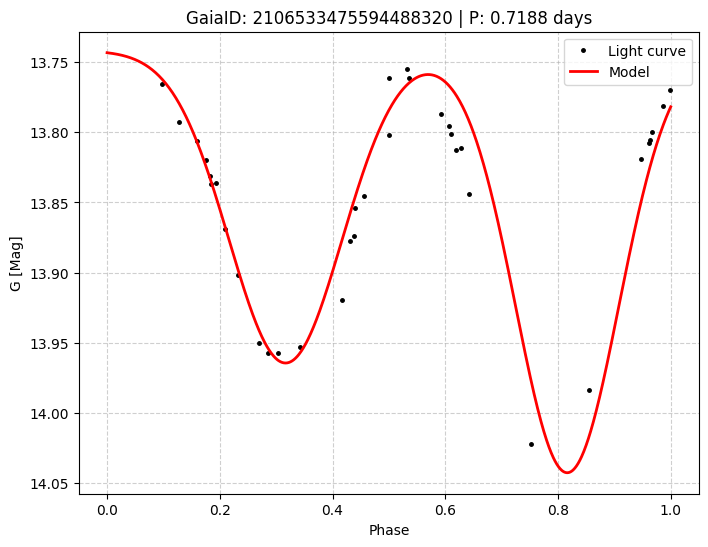}
    \end{minipage}

    \vspace{0.5em}

    \begin{minipage}{0.48\textwidth}
        \includegraphics[width=\linewidth]{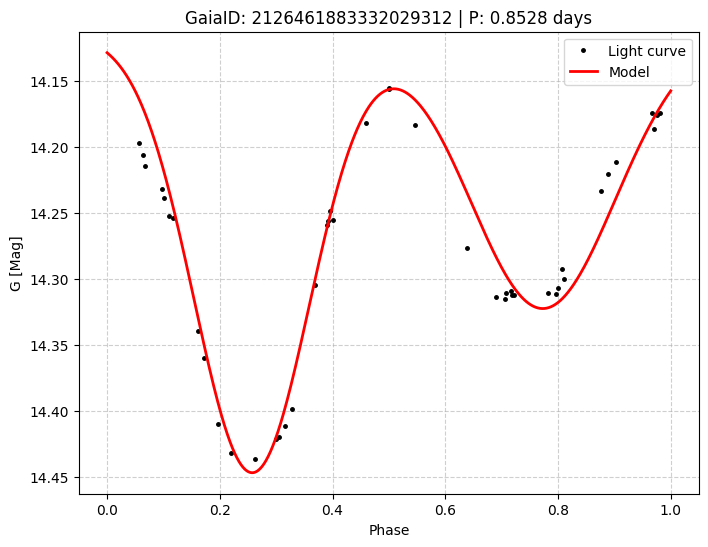}
    \end{minipage} \hfill
    \begin{minipage}{0.48\textwidth}
        \includegraphics[width=\linewidth]{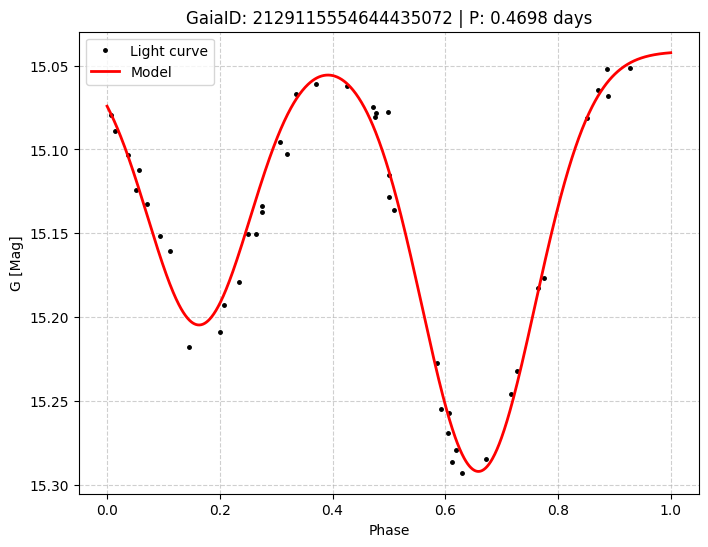}
    \end{minipage}
\end{center}

\subsection{EW samples}
\label{fig:EWsamples}
\begin{center}
    \begin{minipage}{0.48\textwidth}
        \includegraphics[width=\linewidth]{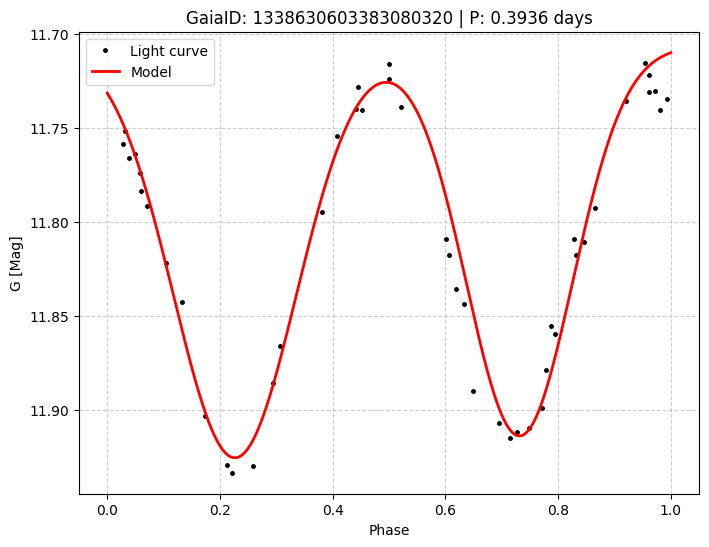}
    \end{minipage} \hfill
    \begin{minipage}{0.48\textwidth}
        \includegraphics[width=\linewidth]{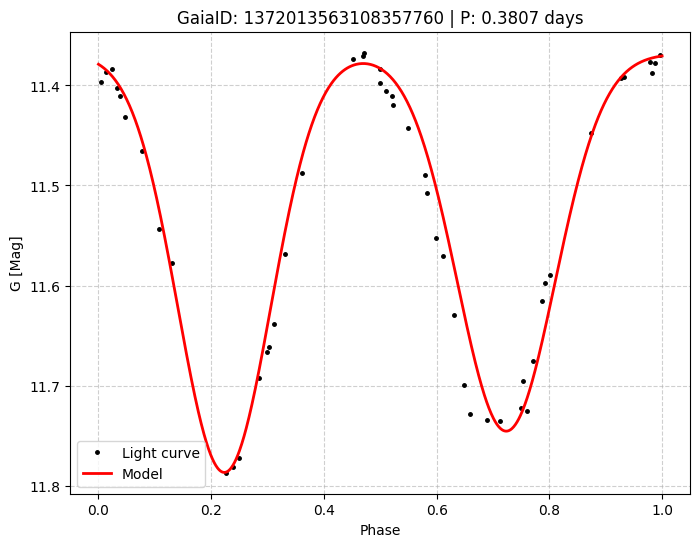}
    \end{minipage}

    \vspace{0.5em}

    \begin{minipage}{0.48\textwidth}
        \includegraphics[width=\linewidth]{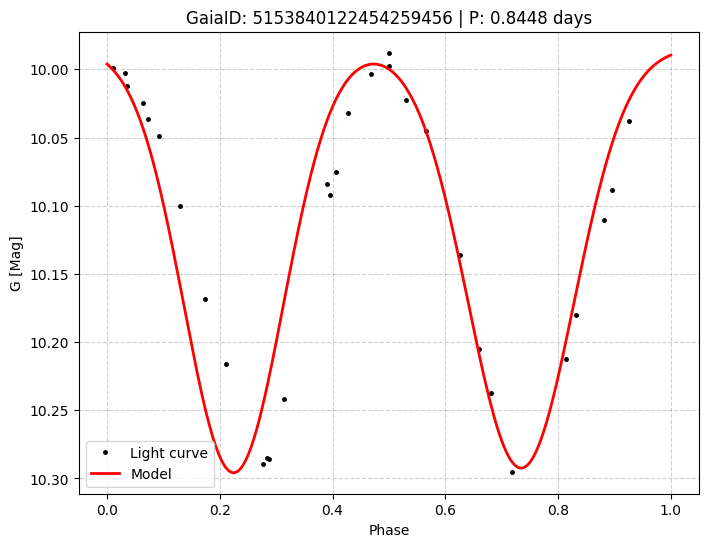}
    \end{minipage} \hfill
    \begin{minipage}{0.48\textwidth}
        \includegraphics[width=\linewidth]{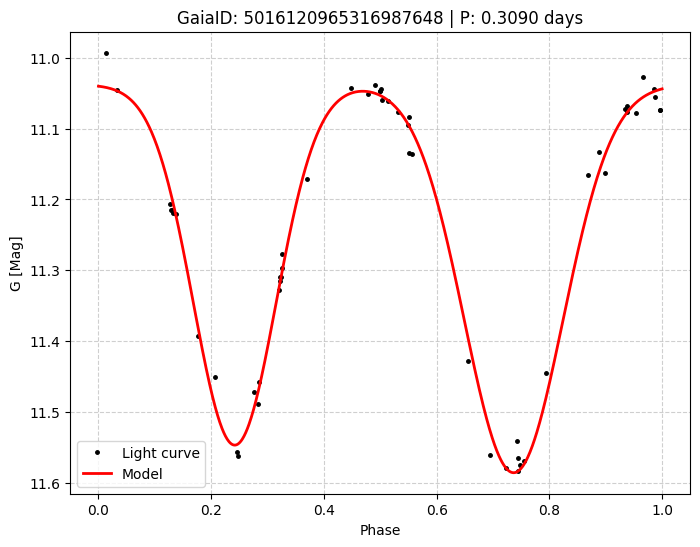}
    \end{minipage}

    \vspace{0.5em}

    \begin{minipage}{0.48\textwidth}
        \includegraphics[width=\linewidth]{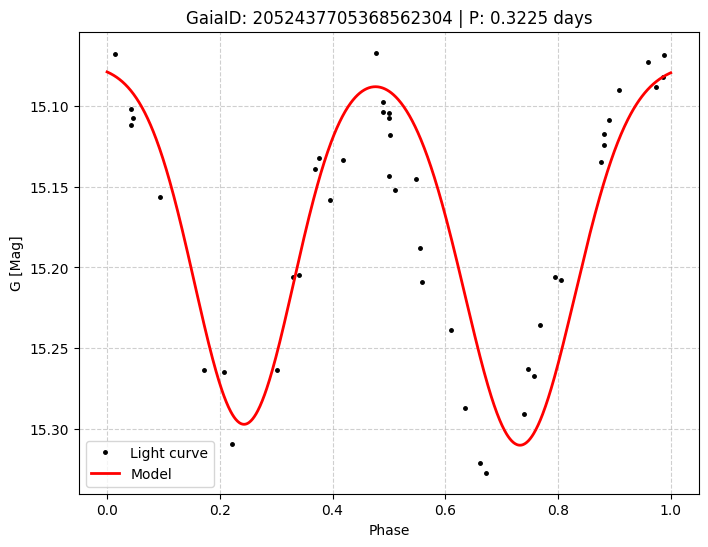}
    \end{minipage} \hfill
    \begin{minipage}{0.48\textwidth}
        \includegraphics[width=\linewidth]{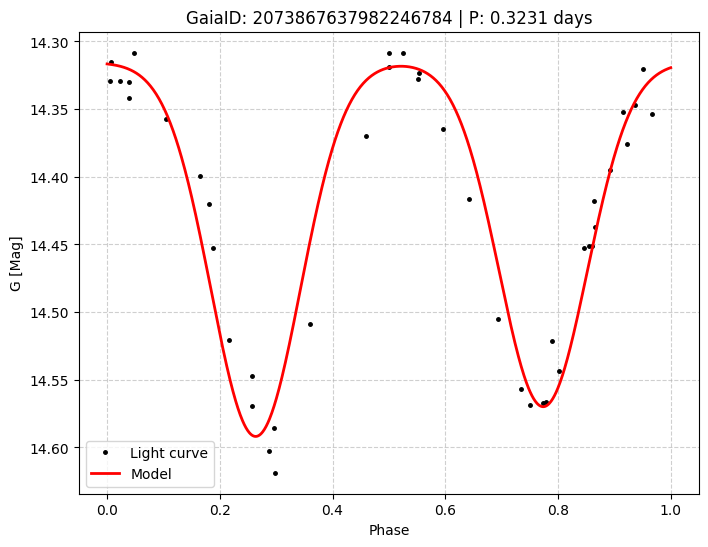}
    \end{minipage}

    \vspace{0.5em}

    \begin{minipage}{0.48\textwidth}
        \includegraphics[width=\linewidth]{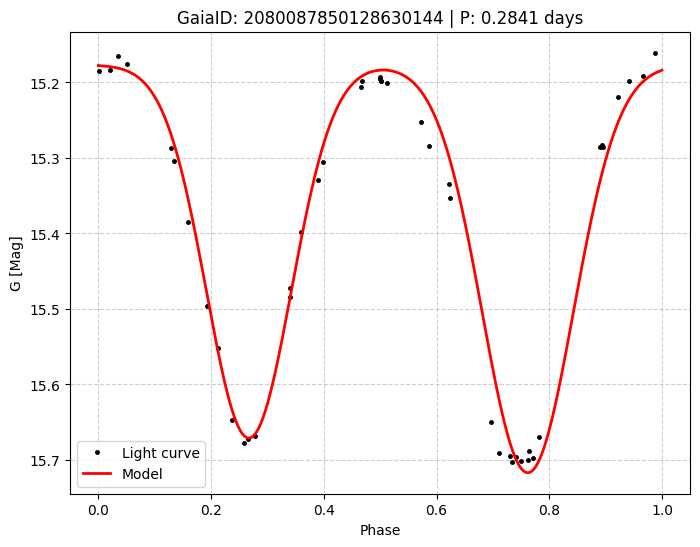}
    \end{minipage} \hfill
    \begin{minipage}{0.48\textwidth}
        \includegraphics[width=\linewidth]{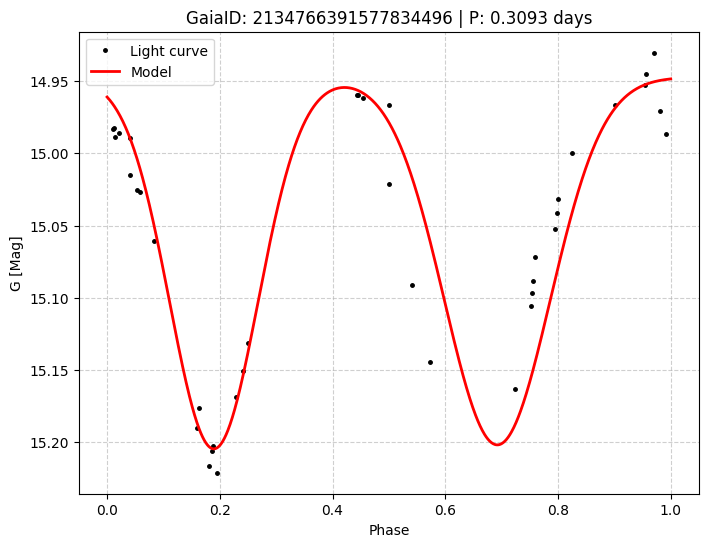}
    \end{minipage}

    \vspace{0.5em}

    \begin{minipage}{0.48\textwidth}
        \includegraphics[width=\linewidth]{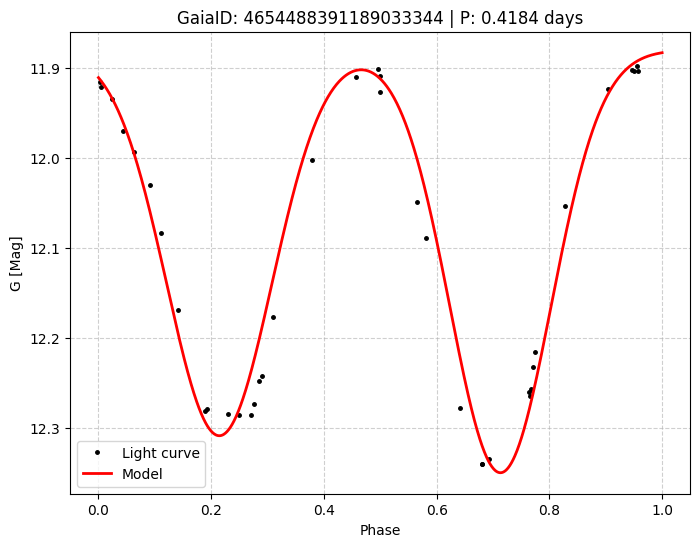}
    \end{minipage} \hfill
    \begin{minipage}{0.48\textwidth}
        \includegraphics[width=\linewidth]{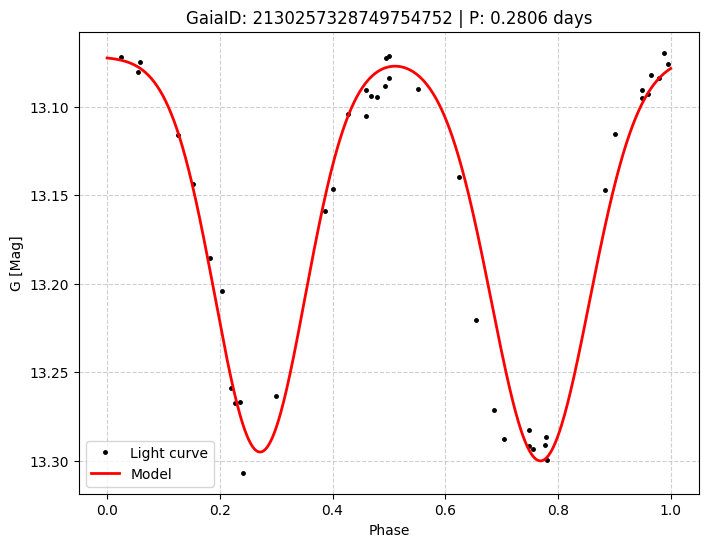}
    \end{minipage}

    \vspace{0.5em}

    \begin{minipage}{0.48\textwidth}
        \includegraphics[width=\linewidth]{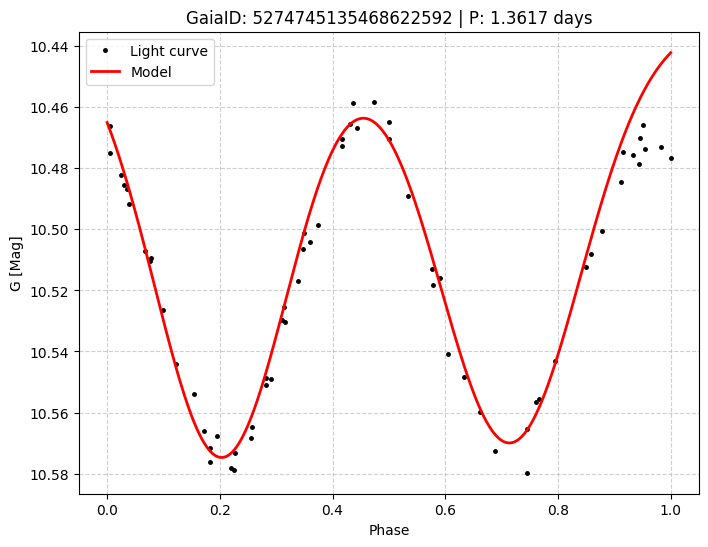}
    \end{minipage} \hfill
    \begin{minipage}{0.48\textwidth}
        \includegraphics[width=\linewidth]{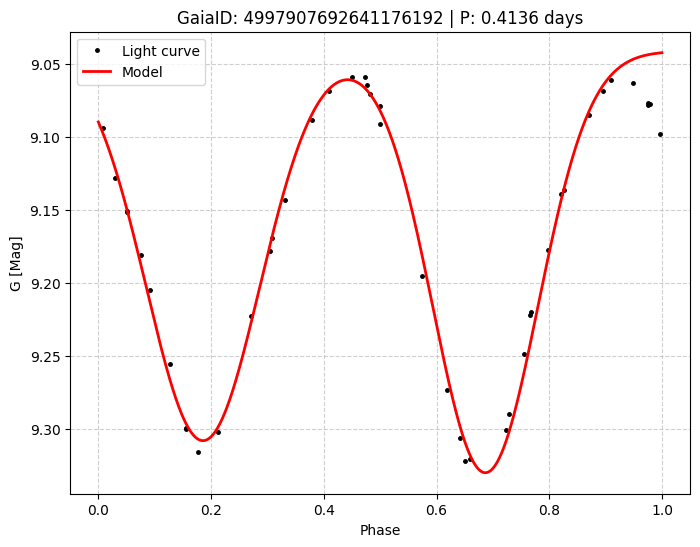}
    \end{minipage}
\end{center}

\twocolumn 

\bibliographystyle{elsarticle-harv} 
\bibliography{example}

\end{document}